\documentclass[prb,twocolumn,showpacs,superscriptaddress]{revtex4-1}

\usepackage{graphicx}
\usepackage{rotating}
\usepackage{amsmath}
\usepackage{bbm}
\usepackage{subfigure}
\usepackage{color}
\usepackage{natbib}
\usepackage{longtable}

\newcommand{\be}{\begin{equation}}
\newcommand{\ee}{\end{equation}}
\newcommand{\bea}{\begin{eqnarray}}
\newcommand{\eea}{\end{eqnarray}}

\begin{document}
\widetext

\title {Self-consistent \textit{GW}: an all-electron implementation with localized basis functions}

\author{Fabio Caruso}
\author{Patrick Rinke}
\affiliation{Fritz-Haber-Institut der Max-Planck-Gesellschaft, Faradayweg 4-6, D-14195 Berlin, Germany} 
\author{Xinguo Ren}
\affiliation{Fritz-Haber-Institut der Max-Planck-Gesellschaft, Faradayweg 4-6, D-14195 Berlin, Germany} 
\affiliation{Key Laboratory of Quantum Information, University of Science and Technology of China, Hefei, 230026, China}
\author{Angel Rubio}
\affiliation{Nano-Bio Spectroscopy group and ETSF Scientific Development Centre, Dpto.\ F\'isica de Materiales, Universidad del Pa\'is Vasco, CFM CSIC-UPV/EHU-MPC and DIPC, Av.\ Tolosa 72, E-20018 San Sebasti\'an, Spain}
\affiliation{Fritz-Haber-Institut der Max-Planck-Gesellschaft, Faradayweg 4-6, D-14195 Berlin, Germany} 
\affiliation{European Theoretical Spectroscopy Facility}
\author{Matthias Scheffler} 
\affiliation{Fritz-Haber-Institut der Max-Planck-Gesellschaft, Faradayweg 4-6, D-14195 Berlin, Germany}

\begin{abstract}
This paper describes an all-electron implementation 
of the self-consistent $GW$ (sc-$GW$) approach -- i.e. 
based on the solution of the Dyson equation --  
in an all-electron numeric atom-centered orbital (NAO) basis set. 
We cast Hedin's equations into a matrix form that 
is suitable for numerical calculations by means of 
i) the resolution of identity technique to 
handle 4-center integrals; and
ii) a basis representation for  
the imaginary-frequency dependence of dynamical operators. 
In contrast to perturbative $G_0W_0$, 
sc-$GW$ provides a consistent framework 
for ground- and excited-state properties and facilitates 
an unbiased assessment of the $GW$ approximation.
For excited-states, we benchmark sc-$GW$ for 
five molecules relevant for organic photovoltaic applications: 
thiophene, benzothiazole, 1,2,5-thiadiazole, naphthalene, and tetrathiafulvalene.
At self-consistency, the quasi-particle energies are found to be in good 
agreement with experiment and, on average, more accurate than $G_0W_0$
based on Hartree-Fock (HF) or density-functional theory with the
Perdew-Burke-Ernzerhof  (PBE) exchange-correlation functional.
Based on the Galitskii-Migdal total energy, structural properties 
are investigated for a set of diatomic molecules.
For binding energies, bond lengths, and 
vibrational frequencies sc-$GW$ 
and $G_0W_0$ achieve a comparable performance, 
which is, however, not as good as that 
of exact-exchange plus correlation in the 
random-phase approximation (EX+cRPA) and its advancement 
to renormalized second-order perturbation theory (rPT2). 
Finally, the improved description of dipole moments 
for a small set of diatomic molecules
demonstrates the quality of the sc-$GW$ ground state density.  
\end{abstract}

\date{\today}
\pacs{}

\maketitle

Many-body perturbation theory (MBPT) \cite{fetter} in the $GW$ approach for
the electron self-energy 
\cite{Hedin1965,thegwmethod,Onida/Reining/Rubio}
provides a natural framework for an {\it ab initio}, parameter-free description of 
photo-ionization processes and charged excitations.\cite{Hedin19701} 
In recent years, the $GW$ approach has become a popular method for the computation
of band gaps and charged excitation energies for extended \cite{patrickpssb,patrick2005} 
and finite systems \cite{Blase/PRB/2011,thygesen}.
In numerical implementations, following Hybertsen and Louie, \cite{hybertsenlouie1986}
it is standard practice to 
treat the $GW$ self-energy as a single-shot perturbation ($G_0W_0$) acting on a 
Kohn-Sham (KS) or Hartree-Fock (HF) reference system. Thus, excitation energies are evaluated
from first-order Feynman-Dyson perturbation theory as corrections
to a set of single-particle eigenvalues.

The popularity of the $G_0W_0$ approximation stems from the substantial 
reduction in the complexity of Hedin's equations at first-order perturbation theory: 
the KS or HF eigenstates from a self-consistent field calculation can be used as
basis functions and provide a convenient representation in which the non-interacting 
Green function is diagonal. 
In this basis, only diagonal matrix elements of the 
self-energy $\Sigma$ are needed to evaluate quasi-particle corrections at first-order.
Thus, $G_0W_0$ grants a considerable simplification of the 
linear algebra operations which is decisive for applying 
the theory to large molecules and solids.

Although numerically more efficient than a non-perturbative approach, 
$G_0W_0$ suffers from several undesirable shortcomings
such as the dependence on the starting point,\cite{patrick2005,Fuchs/etal:2007,benzenepaper,Bruneval/Marques:2013} 
the violation of conservation laws for momentum, total energy and particle number,\cite{baymkadanoff,baym,dahlenleeuwen2006}
and -- most importantly -- the limited access to ground-state properties, 
which are kept unchanged from the preliminary density functional theory (DFT) 
or HF calculations.

It is known that the self-consistent $GW$ approach (sc-$GW$) -- in which both the 
Green function $G$ and the screened Coulomb interaction $W$ are iterated to self-consistency -- 
ameliorates most of the pathologies of perturbative $G_0W_0$.\cite{fabioprl}
A particularly appealing feature of the sc-$GW$ method 
consists in the possibility of treating 
ground- and excited-states {\it at the same level of theory}. 
This property arises by virtue of the non-perturbative nature of the 
sc-$GW$ approach, whereby the Green function is updated and encompasses many-body effects introduced 
by the self-energy. In contrast, in perturbative theories (which generally do not 
introduce updates in the Green function) the electronic structure coincides with 
that of the corresponding starting point.
Therefore, density and total energy -- and derived quantities such as dipole moments, bond lengths, and 
binding energies -- become accessible at self-consistency and reveal the
quality of the $GW$ ground-state. 
Finally, at self-consistency  
excited- and ground-state properties are independent of
the starting point, at least for closed shell systems,\cite{fabioprl} 
and provide an {\it unbiased} assessment of the $GW$ approach.

A previous study on the homogeneous electron gas (HEG) reported 
a deterioration of the sc-$GW$ spectral properties,
as compared to $G_0W_0$.\cite{holmvonbarth1998} 
This has been attributed to a poor description of the satellite 
peaks at self-consistency.\cite{holmvonbarth1998}
For extended systems, the performance of sc-$GW$ remains controversial 
due to the scarce number of calculations for 
real solids.\cite{eguiluz1998,eguiluz2002,kutepov,kutepov2} 
Part of this controversy can be traced back to 
basis set problems in early all-electron calculations\cite{Friedrich/etal:2006} and
to the large influence that pseudo-potentials may have on $GW$ band gaps.\cite{GomezAbal08}
More recently, sc-$GW$ calculations for atoms \cite{stan} 
and molecules \cite{thygesen,fabioprl}  
have shown improvements in the description of the
first ionization energies and for transport 
properties  \cite{PhysRevB.83.115108} of finite systems.

The price to pay in sc-$GW$ is the demanding 
iterative procedure. 
The higher complexity of sc-$GW$ arises for the following reasons. 
i) The Green function obtained from the solution of 
the Dyson equation is in general non-diagonal. This  considerably increases the
computational cost of the evaluation of the dielectric matrix.
ii) The non-diagonal matrix elements 
of $\Sigma$ are needed to solve the Dyson equation.
iii)  Fourier transforms of dynamical quantities are needed that introduce their own computational difficulties. 

In the first part of this paper, we present an all-electron implementation 
of the sc-$GW$ method in the localized 
basis-set code FHI-aims\cite{blum} and propose a recipe to efficiently address points i)-iii).
An optimized set of localized basis functions was used to represent the 
Green function and the self-energy operator.
Non-local two-particle operators, 
such as the screened Coulomb interaction, were computed by means of
the resolution of the 
identity technique\cite{whitten/product/basis,dunlap/product/basis,Mintmire/product/basis} 
(also known as density fitting method) in a general framework 
previously introduced by some of us.\cite{Xinguo/implem_full_author_list}
Finally, an auxiliary basis of Lorentzian functions 
was introduced for an efficient analytical evaluation 
of Fourier transforms between imaginary time and frequency. 

The second part of the paper, focuses on the assessment of ground- and excited-state properties as 
obtained from sc-$GW$ for molecules. 
The quality of the sc-$GW$ ground state was investigated 
by computing binding energies, 
bond lengths, vibrational frequencies, densities, 
and dipole moments for a small set of 
hetero- and homo-atomic dimers.
The full valence excitation spectra were evaluated for a set of molecules 
relevant for organic photovoltaic applications (thiophene, benzoithiazole, 1,2,5-thiadiazole,
naphthalene, and tetrathiafulvalene). 
From this study we conclude that sc-$GW$ systematically 
improves the spectral properties of finite systems over 
the entire excitation spectrum (that is, not only for the first ionization energy) 
as compared to standard perturbative $G_0W_0$ calculations based on semi-local DFT
and HF. Nonetheless, for certain starting points -- exemplified by the PBE0 hybrid 
functional -- $G_0W_0$ slightly outperforms sc-$GW$, providing ionization energies 
in better agreement with experimental reference data, as also previously demonstrated for 
benzene and the azabenzenes in Ref.~\onlinecite{benzenepaper}.
For structural properties the sc-$GW$ method yields a less satisfactory agreement with experiment. 
For dimers, bond lengths and binding energies are slightly underestimated, 
and in this case there is no substantial improvement over perturbative approaches such as 
$G_0W_0$ or the random-phase approximation (RPA).
Finally, self-consistency gives an accurate description of the electron density
as manifested by the accurate dipole moments of diatomic molecules.
These results suggest that sc-$GW$ is a promising method for charge transfer compounds and interfaces. 
However, our study also indicates the importance of 
including higher order exchange and correlation diagrams beyond $GW$ to accurately describe the structural 
properties of molecules.

The paper is organized as follows: 
Section \ref{sec:theory} gives a brief introduction to the $GW$ 
approximation recalling the basic equations needed for the computation of the Green function $G$ and 
the self-energy $\Sigma$.
An optimal representation of Hedin's equations in terms of 
localized basis functions and of the resolution of the 
indentity is presented in Sec.~\ref{sec:repr}.
In Sec.~\ref{sec:ft} we present
the scheme employed in the computation 
of the Fourier integrals of the Green 
function and other dynamical quantities.
We report in Sec.~\ref{sec:spectra} an assessment of sc-$GW$
for the excitation spectra of molecules and,  in Sec.~\ref{sec:migdal}, 
for the ground-state properties of diatomic molecules.
Our conclusions and final remarks are reported in Sec.~\ref{sec:conclusion}.

\section{Theoretical Framework}\label{sec:theory}

In MBPT the complexity of the many-body problem is recast into the
calculation of the single-particle Green function. 
Knowledge of the Green function grants immediate access to the (charged) 
single-particle excitation energies
of the system, to the total energy and, more generally, expectation values of any 
single-particle operator.
Green function theory is well documented in the literature \cite{fetter} and we recall here only the 
basic equations relevant for the $GW$ approach, adhering to Hartree atomic units 
$\hbar = m_e = e^2 = 1$. 

For a system of non-interacting electrons described through a time-independent Hamiltonian, 
the Green function can be written explicitly in terms of the single-particle eigenstates  $\psi_{n}^{\sigma}({\bf r})$ 
and eigenvalues $\epsilon^{\sigma}_{n}$:
\be\label{eq:g}
G_0^{\sigma}({\bf r},{\bf r'},\omega)=\sum_n \frac{\psi_{n}^{\sigma}({\bf r})\psi^{\sigma *}_{n}({\bf r'})}
{\omega-(\epsilon^{\sigma}_{n}-\mu)-i\eta \,sgn(\mu-\epsilon^{\sigma}_{n})}\quad,
\ee
where $n$ and $\sigma$ refer to orbital and spin quantum numbers, respectively.  
$\mu$ is the electron chemical potential, and $\eta$ a positive infinitesimal.
In practice, $\psi^{\sigma}_{n}({\bf r})$ and $\epsilon^{\sigma}_{n}$ 
are generally obtained from the self-consistent 
solution of the Hartree-Fock or (generalized) KS equations. 

For interacting electrons, the Green function has to be evaluated by solving the 
Dyson equation:
\begin{align} \label{eq:dyson}
&G^{\sigma}({\bf r},{\bf r'},\omega) = G_0^{\sigma}({\bf r},{\bf r'},\omega) + \nonumber
\int d{\bf r}_{1}d{\bf r}_{2} G_0^{\sigma}({\bf r},{\bf r}_{1},\omega)\times \\
& \times\left[ \Sigma^{\sigma}({\bf r}_{1},{\bf r}_{2},\omega) + \nonumber
\Delta v_{\rm H}^{\sigma}({\bf r}_{1})\delta({\bf r}_{1}-{\bf r}_{2})-\right. \\
&\left. -v^{\sigma}_{\rm xc}({\bf r}_{1},{\bf r}_{2}) \right] G^{\sigma}({\bf r}_{2},{\bf r'},\omega)\quad.
\end{align}
Here $\Delta v_{\rm H}^{\sigma}$ is the change in the
Hartree potential accounting for density differences 
between $G_0^{\sigma}$ and $G^\sigma$, and
 $v^{\sigma}_{\rm xc}$ is the exchange-correlation part of the single-particle 
Hamiltonian corresponding to the non-interacting Green function $G_0$. 
For example, if $G_0$ is the Hartree-Fock Green function, then $v^{\sigma}_{\rm xc}$
corresponds to the non-local exchange operator $\Sigma_{\rm x}$. 
Alternatively, for a KS Green function, $v^{\sigma}_{\rm xc}$ is the local 
exchange-correlation potential.
The electron self-energy $\Sigma^{\sigma}$ encompasses 
all many-body exchange-correlation effects 
and therefore its practical evaluation requires approximations. 
Following Hedin\cite{Hedin1965,Hedin19701}, the self-energy can be expanded in a 
perturbative series 
of the {\it screened} Coulomb interaction $W$, with the first-order term 
given by the $GW$ approximation:
\be\label{eq:sigma}
\Sigma^{\sigma}({\bf r},{\bf r'},\tau)=i G^{\sigma}({\bf r},{\bf r'},\tau)
W({\bf r},{\bf r'},\tau)\quad.
\ee
By virtue of the time translation invariance it suffices to express $\Sigma$ 
in terms of time differences ($\tau=t-t'$).
The screened interaction $W({\bf r},{\bf r'},\omega)$ is in turn defined through another
Dyson equation:
\begin{align}
W&({\bf r},{\bf r'},\omega) = v({\bf r},{\bf r'}) + \nonumber \\ 
&\int d{\bf r}_1d{\bf r}_2  v({\bf r},{\bf r}_1) \chi({\bf r}_1,{\bf r}_2,\omega) W({\bf r}_2,{\bf r'},\omega)\quad.
\label{eq:W}
\end{align}
Here, $v$ is the bare Coulomb interaction $1/|{\bf r}-{\bf r'}|$ and 
$\chi$ the irreducible polarizability, which in $GW$ is approximated by the 
product of two Green functions:
\be\label{eq:chi0}
\chi({\bf r},{\bf r'},\tau) = -i \sum_\sigma G^{\sigma}({\bf r},{\bf r'},\tau)G^{\sigma}({\bf r'},{\bf r},-\tau)\quad.
\ee

The self-consistent nature of Eqs.~\ref{eq:dyson}-\ref{eq:chi0}
arises from the interdependence of 
the self-energy and the Green function. 
In the $G_0W_0$ approach, the self-energy 
is evaluated non-self-consistently, 
and the Dyson equation is solved approximately in a parturbative fashion. 
The quasi-particle excitation energies are then obtained 
from the quasi-particle equation:
\begin{align}\label{eq:qpe}
\epsilon^{\rm QP}_{n,\sigma}= \epsilon^\sigma_n +{\rm Re}
\langle \psi^\sigma_n| \Sigma^\sigma (\epsilon^{\rm QP}_{n,\sigma})-v_{\rm xc}^\sigma |\psi^\sigma_n \rangle\quad.
\end{align}

In this work,  Eqs.~\ref{eq:dyson}-\ref{eq:chi0} 
are solved fully self-consistently.
In practice, an iterative procedure requires the following steps:
\begin{enumerate} 
\item Construction of an initial non-interacting Green function $G_0$ from a 
preliminary SCF calculation through Eq.~\ref{eq:g}.
\item Evaluation of the polarizability $\chi$ from Eq.~\ref{eq:chi0} and
Fourier transformation of $\chi$ to the frequency domain.
\item Calculation of the screened Coulomb interaction $W$ from Eq.~\ref{eq:W} and
Fourier transformation of $W$ to the time domain.
\item Evaluation of the self-energy $\Sigma$ from the Eq.~\ref{eq:sigma} and 
 Fourier transformation of $\Sigma$ to the frequency domain.
\item Update of the Green function from the Dyson equation (Eq.~\ref{eq:dyson}) and
Fourier transformation of $G$ to the time domain.
\item Mixing of the Green function to accelerate the 
convergence of the self-consistent loop.
\item Iteration of steps 2.-5. until a convergence criterion is satisfied.
\end{enumerate}

In a numerical implementation, a choice for the basis set expansion of the 
quantities in Eqs.~\ref{eq:dyson}-\ref{eq:chi0} has to be made.
Our choice will be discussed in the next Section.


\section{Self-consistent \textit{GW} within a localized basis}\label{sec:repr}

Previous implementations of sc-$GW$ were based on 
Gaussians or Slater orbitals,\cite{holmvonbarth1998,stan}
full potential linear augmented plane-waves,\cite{eguiluz2002,kutepov}
real-space grids, \cite{Pablo:2002}
and numeric atom-centered orbitals (NAO).\cite{thygesen}
In the present work, the Green function $G$, the self-energy $\Sigma$, and all single-particle operators,
are expanded in a numeric atom-centered orbital basis $\lbrace\varphi_i ({\bf r})\rbrace$, 
with basis functions of the form:
\be
 \varphi_i({\bf r}) = \frac{u_i(r)}{r}Y_{lm}(\Omega)\quad,
\ee
where ${u_i(r)}$ are numerically tabulated radial functions 
and $Y_{lm}(\Omega)$ spherical harmonics. 
For numerical convenience, we work with real-valued basis functions
by requiring -- without loss of generality -- that
$Y_{lm}(\Omega)$ denotes the real part (for $m=0,\dots,l$) and the imaginary part  (for $m=-l,\dots,-1$) of complex
spherical harmonics.
In FHI-aims the choice of the radial functions ${u_i(r)}$ is not limited. In this work we will show results for numerically tabulated Gaussian orbital basis sets and the \emph{Tier} hierarchy of FHI-aims for NAOs.\cite{blum}
We refer to Ref.~\onlinecite{blum} for details on the construction and optimization as well as the properties of the NAO basis sets in FHI-aims.

In terms of the $\varphi_i$ basis functions, the  Green function can be expanded as:
\be\label{EQ:G2}
G^\sigma({\bf r},{\bf r'},i\omega) = \sum^{N_{\rm basis}}_{ijlm}\varphi_i({\bf r})s_{ij}^{-1}G^\sigma_{jl}
(i\omega)s_{lm}^{-1} \varphi_m({\bf r'})\quad,
\ee
where $s_{ij}= \int d{\bf r} \varphi_i({\bf r})\varphi_j({\bf r}) $ is the overlap matrix taking into 
account the non-orthonormality of the basis set and ${N_{\rm basis}}$ is the 
total number of basis functions. In the following, sums over latin indixes $i,j,l,m$ are 
implicitly assumed to run from $1$ to ${N_{\rm basis}}$, whereas sums over $n$ run 
over the total number of states.
The coefficients $G^\sigma_{ij}(i\omega)$ of the expansion are given by:
\be \label{EQ:G3}
G^\sigma_{ij}(i\omega) = \int d{\bf r}d{\bf r'} 
\varphi_i({\bf r}) G^\sigma({\bf r},{\bf r'},i\omega) \varphi_j({\bf r'})\quad.
\ee
The representation in Eq.~\ref{EQ:G3} can be easily applied to 
the non-interacting Green function in Eq.~\ref{eq:g}, yielding
\begin{align}\label{eq:G4}
G^\sigma_{0,ij}(i\omega)&= \sum_{n} \sum_{lm} \frac{s_{il}c^\sigma_{ln}c^\sigma_{mn}s_{mj}}
{i\omega-(\epsilon^{\sigma}_{n}-\mu)}\quad,
\end{align}
where we introduced the expansion of the HF/KS orbitals in the NAO basis 
$\psi^\sigma_{n}({\bf r}) = \sum_l c^\sigma_{ln}\varphi_l({\bf r})$ and 
the Green function was continued to the imaginary frequency axis.
The matrix representation in Eq.~\ref{EQ:G3} is also adopted for the
Hartree potential $v^\sigma_{\rm H}$, the self-energy $\Sigma^\sigma$, 
and the exchange-correlation potential $v^\sigma_{\rm xc}$.
We emphasize that in our implementation the summation over empty states
-- which is at the origin of the basis-set convergence problem of $GW$ calculations 
\cite{PhysRevB.79.201104/product/basis,PhysRevB.78.085125,PhysRevB.82.041103,PhysRevB.81.115105} --
enters only through Eq.~\ref{eq:G4}. 
The self-energy and the polarizability are evaluated as functionals of the Green function, 
and therefore do not any involve any explicit empty-state summation.

To rewrite Hedin's equations in a matrix form suitable for a numerical implementation, 
we need to introduce a matrix representation for two-particle operators.
The expansion of two-particle operators in a numerical 
basis, cannot be handled efficiently through Eq.~\ref{EQ:G3}
due to the appearance of the 4-orbital 2-electron Coulomb integrals of the form:
\begin{align} \label{eq:4centers}
(ij|kl)&=\int\frac{\varphi_{i}({\bf r})\varphi_{j}({\bf r})\varphi_{k}({\bf r'})
\varphi_{l}({\bf r'})}{|{\bf r}-{\bf r'}|}d{\bf r}d{\bf r'}\quad .
\end{align}
The computation of the Coulomb repulsion integrals 
in Eq.~\ref{eq:4centers} is a problem extensively discussed in 
the literature 
\cite{Feyereisen1993359/product/basis,Weigend1998143/product/basis,B204199P/product/basis,eshuis:234114/product/basis,PhysRevB.79.201104/product/basis,Friedrich2009347/product/basis,PhysRevB.49.16214/product/basis,B204199P/product/basis,Weigend1998143/product/basis,Feyereisen1993359/product/basis,qpscgw2004,qpscgw2006} 
and efficient techniques have been developed over the years to make this 
calculation affordable. 
Numerically, the difficulty arises from the large number of NAO pairs and  
from the memory requirements of storing the 4-index matrix $(ij|kl)$.
In the NAO framework, the integrals in Eq.~\ref{eq:4centers} are often 
evaluated by introducing an auxiliary basis set $\lbrace P_\mu(\bf r)\rbrace$, 
with basis functions $P_\mu(\bf r)$ defined to 
span the Hilbert space of  NAO pairs 
\be\label{eq:pairs}
\varphi_i({\bf r})\varphi_j({\bf r} ) \simeq \sum_{\mu=1} C^\mu_{ij} P_\mu({\bf r}) \quad ,
\ee
where $C^\mu_{ij}$ are the coefficients of the expansion. 
Due to the high linear dependence of the NAO products, the number of product basis functions ${N_{\rm aux}}$ 
is much smaller than the number of NAO pairs $O(N_{\rm basis}^2)$, 
making the numerical evaluation of Eq.~\ref{eq:4centers} affordable.
This techique, known as the {\it resolution of the identity} (RI) -- or also 
density-fitting technique -- 
was implemented in the FHI-aims code and we refer to Ref.~\onlinecite{Xinguo/implem_full_author_list} 
for a detailed account of the variational approach employed in the 
determination of the RI coefficients $C^\mu_{ij}$ and for a review of the 
overall accuracy of the RI approach for correlated calculations.

In short, we used the ``RI-V'' variant of the RI scheme, in which the expansion coefficients
are given by: 
\begin{align}
C^\mu_{ij}=\sum_\nu \left( ij | \nu \right) V^{-1}_{\nu\mu}\quad,
\end{align}
where $\left( ij | \nu \right) \equiv\int d{\bf r} \varphi_i({\bf r})\varphi_j({\bf r} )  P_\mu({\bf r'})/|{\bf r}-{\bf r'}|$
and $V_{\nu\mu}$ denotes matrix elements of the Coulomb matrix in the auxiliary basis, i.e., 
$V_{\mu\nu}=\int  d{\bf r}  d{\bf r'} P_\mu({\bf r}) P_\nu({\bf r'}) / |{\bf r}-{\bf r'}|$.
For numerical efficiency, it is convenient to work with the generalized coefficients:
\begin{align}\label{eq:mcoeff}
M^\mu_{ij}=\sum_{\nu} C^{\nu}_{ij} V^{1/2}_{\nu\mu}\quad.
\end{align}
Following Ref.~\onlinecite{Xinguo/implem_full_author_list}, one can write the RI
version of the Dyson equation for the screened Coulomb interaction (Eq.~\ref{eq:W}) as:
\begin {align}
\overline W_{\mu\nu} (i\omega)\equiv  \left [v^{-1}W (i\omega) \right]_{\mu\nu} 
&= \left [ 1 - \Pi(i\omega) \right]^{-1}_{\mu\nu}\quad ,\label{eq:W2}
\end{align}
where we defined $\left[ \Pi(i\omega)\right]_{\mu\nu} \equiv \left[ \chi(i\omega)v \right]_{\mu\nu}$.
In contrast to the RI-based implementation of the $G_0W_0$ method\cite{Xinguo/implem_full_author_list}, 
the operator $\Pi$ has to be expressed as an explicit functional of $G$. Moreover, all non-diagonal
matrix elements in the Green function have to be included. 
These two criteria are satisfied by evaluating $\Pi$ in terms of the $M^\mu_{ij}$ coefficients, as:
\begin {align}
\left[\Pi(i\tau)\right]_{\mu\nu} 
&=-i \sum_\sigma \sum_{ijlm} M_{il}^\mu M_{jm}^\nu \overline G^{\sigma}_{ij}(i\tau)
\overline G^{\sigma}_{lm}(-i\tau).\label{eq:chi2}
\end{align}
Here we defined 
\begin{equation}
\overline G_{ij}(i\tau) \equiv \sum_{lm} s^{-1}_{il} G_{lm}(i\tau)s^{-1}_{mj} \quad .
\end{equation}
The quadruple sum in Eq.~\ref{eq:chi2} may be reduced to  
double sums -- with a considerable reduction of computational cost --  by introducing the intermediate quantity
$A^\mu_{lj,\sigma}(i\tau) \equiv \sum_i M^\mu_{il}\overline G^\sigma_{ij}(i\tau)$. 
In terms of these coefficients Eq.~\ref{eq:chi2} becomes: 
\begin {align}
\left[\Pi(i\tau)\right]_{\mu\nu} 
&=-i \sum_\sigma \sum_{lj} A^\mu_{lj,\sigma}(i\tau) A^\nu_{jl,\sigma}(-i\tau)\quad. 
\end{align}

The self-energy can be evaluated in terms of Eq.~\ref{eq:W2}
providing the following matrix representation of Eq.~\ref{eq:sigma}: 
\begin {align}\label{eq:sigma2}
\Sigma^{\sigma}_{ij}(i\tau) = \frac{i}{2\pi}\sum_{lm}\sum_{\mu\nu} 
M^{\mu}_{il}M^{\nu}_{jm}\overline G^{\sigma}_{lm}(i\tau)\overline W_{\mu\nu}(i\tau)\quad.
\end{align}
By introducing the auxiliary quantity $B^\mu_{jm}(i\tau)=\sum_\nu M^\nu_{jm}\overline W_{\mu\nu}(i\tau)$,
the self-energy can again be cast into a double-sum form:
\begin {align}\label{eq:sigma3}
\Sigma^{\sigma}_{ij}(i\tau) = \frac{i}{2\pi}\sum_{m}\sum_{\mu} 
A^\mu_{im,\sigma}(i\tau)B^\mu_{jm}(i\tau)\quad.
\end{align}
The correlation (exchange) contribution to the self-energy 
can be derived straightforwardly from Eq.~\ref{eq:sigma2} 
by substituting $\overline W$ with 
${\overline W}^{\rm c}_{\mu\nu}\equiv \overline W_{\mu\nu} - \delta_{\mu\nu}$ 
($ {\overline W}^{\rm x}_{\mu\nu}\equiv \delta_{\mu\nu}$).
The Hartree potential $v^\sigma_{\rm H}$ is also 
evaluated as an explicit functional of $G$ as: 
\begin {align}\label{eq:Hartree}
v_{{\rm H},ij}^{\sigma} = \sum_{lm}\sum_{\mu} 
M^{\mu}_{ij}M^{\mu}_{lm}\overline G^{\sigma}_{lm}(i\tau=0^-)\quad.
\end{align}
Finally, the matrix representation of the Dyson equation for the 
Green function completes the set of Hedin's equations:
\begin {align}\label{Dyson}
\overline G_{ij}^{\sigma}(i\omega)= 
\left[\overline G_{0}^{\sigma}(i\omega)^{-1}- 
\Sigma^\sigma(i\omega)+v^\sigma_{\rm xc}-\Delta v^\sigma_{\rm H} \right]_{ij}^{-1}. 
\end{align}
Here $\Delta v^\sigma_{\rm H}$ is the difference between the Hartree potential of 
the interacting and the non-interacting Green function.

\begin{figure}
\begin{center}
\includegraphics[width=0.48\textwidth]{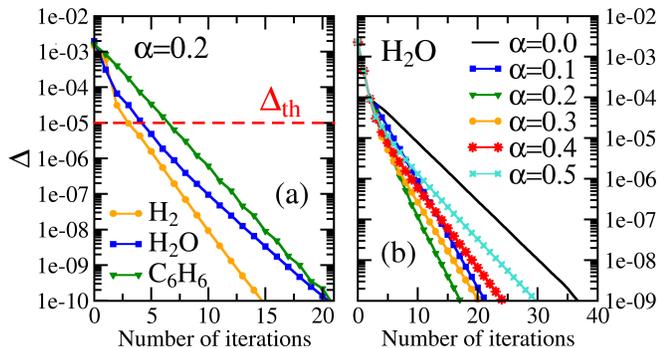}
\caption{\label{fig:loop/conv}
(Color online)
Panel (a): values of $\Delta$ -- defined in Eq.~\ref{eq:delta/conv} --
as a function of the number of iterations of the sc-$GW$ loop
for  H$_2$, H$_2$O and C$_6$H$_6$ in their equilibrium geometry in a Tier 2 basis set.
A linear mixing parameter $\alpha=0.2$ was used for all molecules.
$\Delta_{\rm th}$ indicates the default value of the convergence threshold.
Panel (b): values of $\Delta$ for  H$_2$O as a function of the number of iterations
for different values of $\alpha$.
}
\end{center}
\end{figure}

To facilitate the convergence of the sc-$GW$ loop, the input Green
function $\overline G^{\rm input}$ of the $(N+1)$-th iteration
is obtained from a linear mixing scheme:
\begin{align}
\overline G_{ij}^{\rm input} (i\tau) = 
\alpha \overline G_{ij}^{N} (i\tau) + (1-\alpha) \overline G_{ij}^{N-1} (i\tau)\quad,
\end{align}
where $\overline G ^N$ denotes the Green function obtained from the $N$-th solution of the Dyson equation,
and $\alpha$ is the mixing parameter.
As illustrated in panel (b) of Fig.~\ref{fig:loop/conv}, we find that $\alpha=0.2$ is typically a good choice. 
The convergence of the self-consistent loop is monitored looking at the average
deviation of the Green function at each iteration $\Delta$, defined as:
\begin{align}\label{eq:delta/conv}
\Delta = \frac{1}{N_{\rm basis}^2}\sum_{i,j}|\overline G^N_{ij}(i\tau=0^-) -\overline G^{N-1}_{ij}(i\tau=0^-)|\quad.
\end{align}
The sc-$GW$ loop is considered converged when $\Delta$ drops below a
chosen threshold $\Delta_{\rm th}$. Default settings used in most
calculations are $\Delta_{\rm th}=10^{-5}$, which suffices to converge
both total and quasi-particle energies.
The convergence of sc-$GW$ is illustrated in Fig.~\ref{fig:loop/conv},
where $\Delta$ is reported as a function of the number of
iterations for H$_2$, H$_2$O and C$_6$H$_6$.

Equations \ref{eq:W2}-\ref{Dyson} constitute a matrix representation of 
Hedin's equations in the $GW$ approach (Eqs.~\ref{eq:dyson}-\ref{eq:chi0}). 
We emphasize again that in Eqs.~\ref{eq:W2}-\ref{Dyson}: 
i) all electrons are treated on the same quantum mechanical level, i.e. fully self-consistently; 
ii) no model screening was used in the calculation of $W$;  
iii) all non-diagonal matrix elements of $G$ and $\Sigma$ are correctly accounted for.

\begin{figure}
\includegraphics[width=0.4\textwidth]{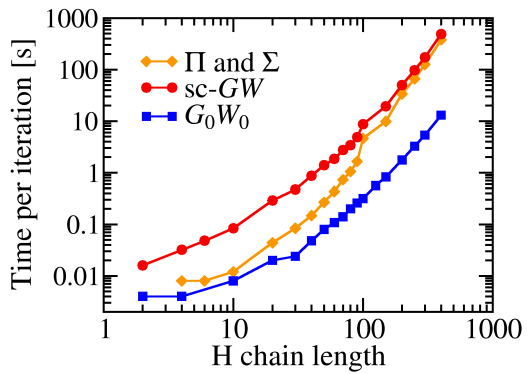}
\caption{\label{fig:scaling} 
(Color online) Total time (in seconds, on a single CPU) per iteration of the sc-$GW$ loop, for linear 
hydrogen chains of different lengths. The total time required for the evaluation of the 
self-energy in $G_0W_0$ is included for comparison.
}
\end{figure}

The evaluation of Eqs.~\ref{eq:chi2} and \ref{eq:sigma2} 
is the most computationally demanding operation of our implementation. 
The scaling of the computational time as a function of the
basis set size equals that of $G_0W_0$ calculations but with a larger prefactor. 
To illustrate this aspect, we report in Fig.~\ref{fig:scaling} the total computational time 
spent for a single iteration of Eqs.~\ref{eq:W2}-\ref{Dyson} as function of the length of a 
linear hydrogen chain in a minimal basis set (i.e., with one NAO per atom).
As compared to conventional $G_0W_0$ implementations, 
the additional computational cost arises from the 
necessity of accounting for non-diagonal 
matrix elements in the calculation of $G$ and $\Sigma$.

\begin{figure}
\begin{center}
\includegraphics[width=0.48\textwidth]{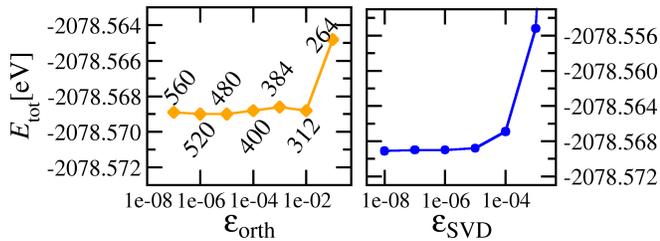}
\caption{\label{fig:RI_conv}
sc-$GW$ total energy of H$_2$O as a function of the 
convergence parameters $\varepsilon_{\rm orth}$ (left panel) and $\varepsilon_{\rm SVD}$ (right panel), 
evaluated with a Tier 2 basis set. 
The number of product basis functions corresponding 
to each value of $\varepsilon_{\rm orth}$ is also reported.
}
\end{center}
\end{figure}

The only approximation introduced up to 
this point is the resolution of the identity for 
the expansion of the product of NAO pairs  (Eq.~\ref{eq:pairs}).
As discussed in Ref.~\onlinecite{Xinguo/implem_full_author_list}, the accuracy of the RI can be 
monitored systematically by means of two control parameters: 
$\varepsilon_{\rm orth}$ and $\varepsilon_{\rm SVD}$. 
$\varepsilon_{\rm orth}$ sets the accuracy threshold 
for the Gram-Schmidt orthonormalization 
employed for the reduction of the linear dependence of 
on-site (i.e. on the same atom) product basis functions $P_\mu$.
In practice, by chosing smaller values of $\varepsilon_{\rm orth}$ 
one may increase the number of product basis functions used in the expansion in Eq.~\ref{eq:pairs}. 
Similarly, the parameter $\varepsilon_{\rm SVD}$ controls the 
singular value decomposition (SVD) for the
orthonormalization of product basis functions on different atoms.
A more detail description of the effects of these 
parameters can be found in Ref.~\onlinecite{Xinguo/implem_full_author_list}.
To show the effect of the RI scheme on the self-consistent Green function, 
we report in Fig.~\ref{fig:RI_conv} the  sc-$GW$ total energy --
evaluated from Eq.~\ref{eq:rewriting}, introduced in Sec.~\ref{sec:migdal} --  of
the water molecule as a function of $\varepsilon_{\rm orth}$ 
(left panel), and $\varepsilon_{\rm SVD}$ (right panel). 
For a wide range of values of the control parameters 
$\varepsilon_{\rm orth}$  and $\varepsilon_{\rm SVD}$,
the changes of the total energy are of the order of $10^{-4}$ eV or less.
In all following calculations we therefore used  $\varepsilon_{\rm orth}=10^{-5}$ and 
$\varepsilon_{\rm SVD}=10^{-5}$.


\section {Discretization of the Fourier integrals}\label{sec:ft}

\begin{figure}
\includegraphics[width=0.4\textwidth]{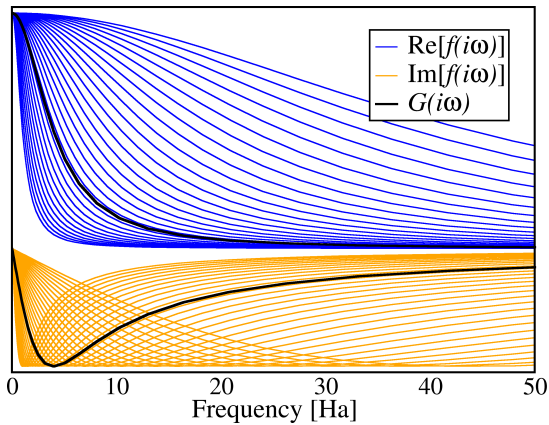}
\caption{\label{fig:pole} 
(Color online) Comparison of the analytic structure of the real 
(in blue, above) and imaginary part (in orange, below) 
of $f_n(i \omega)$ for different values of $b_n$, 
and a matrix element of the Green function (black).}
\end{figure}

In this implementation, we solve Eqs.~\ref{eq:W2}-\ref{Dyson} in imaginary 
time and frequency, taking advantage of the reduced number 
of frequency points required to describe 
$G^{\sigma}(i\omega)$ and other dynamical quantities, 
as compared to real frequency implementations.
In a mixed time-frequency formalism
convolutions on the frequency axis can be 
expressed, by virtue of the convolution theorem, 
as products on the time axis after a Fourier transform. 
Due to the slow decay of $G^\sigma(i\omega)$ at large frequencies,
the computation of Fourier transforms
may require extended and dense frequency grids.
We obviate this problem by introducing a basis for the 
frequency/time dependence of all dynamical quantities. 
This permits an analytic evaluation of the Fourier integrals -- as 
one can choose basis functions with a Fourier transform known analytically -- and 
substantially reduces the number of 
frequency points needed to converge the calculation.

Following the approach introduced in Ref.~\onlinecite{Held/ft}, we expand 
the Green function in a set of Lorentzian functions of the form
$f_n(i\omega) = 1/(b_n+i\omega)$, with Fourier transform
 ${f}(i\tau) = 1/(2\pi) e^{-b_n\tau}$.
The parameters $b_n$ are fixed at the beginning of 
the calculation and are distributed logarithmically 
in the energy range covered by the Kohn-Sham or Hartree-Fock eigenvalues $\epsilon^\sigma_n$. 
\begin{figure}
\includegraphics[width=0.45\textwidth]{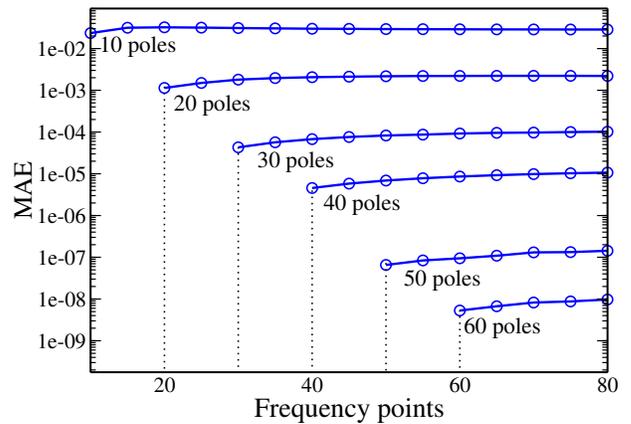}
\caption{\label{fig:FT_acc} (Color online) Mean absolute error (MAE) introduced by 
Fourier transforming the Green function of N$_2$
for different numbers of poles and frequency points.} 
\end{figure}
Although in principle other functions could be used, the functions $f_n(i\omega)$ 
constitute a natural choice for the expansion of the Green function, as the frequency dependence 
of $f_n(i\omega)$ closely resembles the analytic structure of $G$ and 
captures the $1/i\omega$ behaviour at large frequency. 
This is illustrated in Fig.~\ref{fig:pole}, where
the real and imaginary parts of $f_n(i\omega)$ 
-- with different values of $b_n$ -- are compared to a Green function matrix element 
for the Ne atom.
The Green function can be expanded in the basis of Lorentzian functions as:
\be\label{G_re}
G_{ij}^\sigma(i\omega) = \sum_{n=1}^{N_{\rm poles}}
\left[\alpha^n_{ij} f_n^{\rm Re}(i\omega) 
+  \beta^n_{ij} f_n^{\rm Im}(i\omega)
\right],
\ee
where $N_{\rm poles}$ denotes the number of functions $f_n$, 
and $f_n^{\rm Re\, (Im)}(i\omega)$ the real (imaginary) part of $f_n(i\omega)$.
The real and imaginary part of the Green function 
have been treated separately 
to maintain a real-valued
linear-least square problem, leading in turn to
real-valued coefficients $\alpha_n$ and $\beta_n$.
Since the Fourier transform of the $f_n(i\omega)$ 
is known, 
the coefficients $\alpha_n$ and $\beta_n$ also determine the 
expansion of the Green function in imaginary time.
Expansions similar to Eq.~\ref{G_re} were employed also for 
the Fourier transform of $\chi$, $W$ and $\Sigma$.

The imaginary time and frequency axes are then discretized 
on exponentially spaced grids
composed of $N_{\omega}$ 
points in the range $\lbrace0,\omega_{\rm max}\rbrace$, 
and by $2N_{\tau}+1$ points in the range
$\lbrace-\tau_{\rm max},\tau_{\rm max}\rbrace$. The grid points $\omega_n$ 
and integration weights $w(\omega_n)$ are defined as: 
\begin{align}\label{grid}
\begin{array}{ccc}
\omega_k= \omega_0 \left[e^{(k-1)h} -1\right], & &
w(\omega_k) = h\omega_0 e^{(k-1)h} 
\end{array}
\end{align}
and similarly for $\tau_k$ and $w(\tau_k)$.
The constant $h$ is obtained by imposing the maximum frequency $\omega_{\rm max}$ from the constraint 
$\omega_{\rm max} = \omega_0\left[ e^{N_{\omega}h} -1 \right]$
and the parameter $ \omega_0$ sets the initial spacing of the grid.
Typical values adopted in our calculations are 
$\omega_{\rm max}=5000\,\, {\rm Ha}$, $\tau_{\rm max}=1000\,\, {\rm Ha}^{-1}$ and $\omega_0 = \tau_0 = 0.001$.

The error introduced by the Fourier transform can 
be quantified for functions known analytically on both the (imaginary) frequency and time axes such as, 
for instance, the non-interacting Green function given in Eq.~\ref{eq:g}.
In Fig.~\ref{fig:FT_acc}, we report the mean absolute error (MAE)
in the Fourier transform of the non-interacting Green function $G^{\sigma}_0(i\omega)$ 
of the nitrogen dimer N$_2$, averaged over all matrix elements.
The MAE drops exponentially when increasing the number of functions $f_n$, 
and few tens of frequency points suffice to converge the 
Fourier integrals with an accuracy of the order
of $10^{-8}$. In our calculations we used $N_{\omega}=N_{\tau}=60$ as default parameters. 


\begin{figure*}
\includegraphics[width=0.9\textwidth]{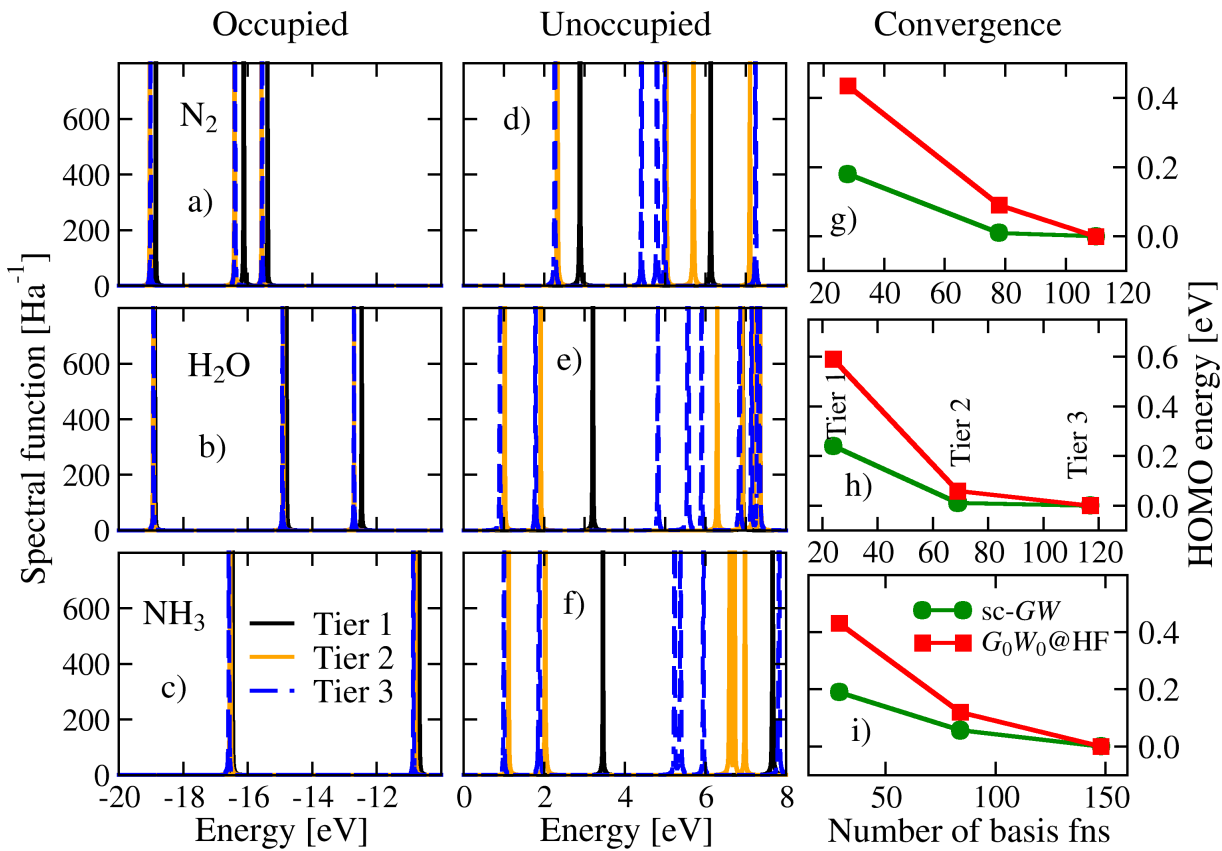}
\caption{\label{fig:spectrum} (Color online) 
Spectral function calculated from Eq.~\ref{eq:spectrum} 
using a sc-$GW$ Green function for H$_2$O, NH$_3$ and
N$_2$ with Tier 1 (black), Tier 2 (orange) and Tier 3 (blue, dashed) NAO basis sets. 
Panels a), b) and c) show the peaks corresponding 
to the first valence states, peaks relatives to conduction states 
are reported in panels d), e) and f). 
Finally, the panels g), h), and i) report the convergence of the HOMO level as a function of the 
number of numerical orbitals used in the basis set.} 
\end{figure*}

\section{Spectral properties of molecules}\label{sec:spectra}

We turn now to the spectral properties in sc-$GW$.
At self-consistency, the excitation spectrum is given by 
the spectral function:
\begin{align}
A(\omega)&=-1/\pi\int d{\bf r}\lim_{{\bf r'}\rightarrow{\bf r}}{\rm Im}\,\, G({\bf r},{\bf r'},\omega)\nonumber \\
 &= -1/\pi\,\, {\rm Tr} \left[  {\rm Im}\,\, G(\omega) \right] \quad \label{eq:spectrum},
\end{align}
where the Green function has to be evaluated on the real frequency axis.
To evaluate  Eq.~\ref{eq:spectrum}, 
we first obtain the real frequency self-energy by means 
of the analytic continuation based on a two-pole fitting scheme \cite{Rieger1999211}.
In this approach, the matrix elements of the self-energy 
in the imaginary frequency domain (i.e., the Fourier transform of 
Eq.~\ref{eq:sigma3}) are fit by polynomials of the form:
\begin{align}\label{eq:pade}
\Sigma(i\omega)&\simeq \sum_{n=1}^{2}
\frac{a_n }{i\omega+b_n}\quad .
\end{align}
Here the matrix element indices were 
suppressed for notational simplicity and the coefficients $a_n$ and $b_n$ are determined by means of
a non-linear least-square fit, solved with a Levenberg-Marquardt algorithm. 
By replacing $i\omega$ by $\omega$ in Eq.~\ref{eq:pade} the self-energy can then be 
evaluated on the real frequency axis.
Once the real-frequency self-energy is obtained, 
the Dyson equation  
is solved directly in real frequency 
on a fine, equally spaced grid.
The resulting Green function is used to determine the sc-$GW$ spectral function $A(\omega)$. 

Previous works\cite{Rieger1999211} have indicated that the
two-pole model presented in Eq.~\ref{eq:pade}
reliably reproduce quasi-particle energies with an average relative error of the $0.2\%$ for solids.
The parameter $\eta$ in the denominator 
of  Eq.~\ref{eq:g}, necessary to avoid the numerical 
divergence of $G_0$ 
is set to $\eta=10^{-4}$. This parameter contributes negligibly 
to the broadening of the spectral function and has no effect on the 
position of the quasi-particle peaks.

As an example, we report the sc-$GW$ spectral functions of  H$_2$O, NH$_3$ and
N$_2$ in Fig.~\ref{fig:spectrum} calculated using basis sets of increasing size.
The sc-$GW$ spectral function shows sharp $\delta$-function-like peaks at the quasi-particle energies. 
The absence of broadening in the 
quasi-particle peaks in Fig.~\ref{fig:spectrum} 
may be associated with a infinite lifetime of the 
corresponding quasi-particle states, as expected for 
states close to the Fermi energy. 
As discussed in Sec.~\ref{sec:life}, 
higher energy excitations may decay through 
the formation of electron-hole pairs, leading to 
a finite lifetime and thus to a 
more pronounced broadening of the quasi-particle peaks.
In panels a), b) and c) in Fig.~\ref{fig:spectrum} we report 
the spectral function corresponding to the highest occupied quasi-particle states
evaluated with a Tier 1, Tier 2 and Tier 3 basis; panels d), e) and f) 
show the peaks corresponding to the lowest unoccupied quasi-particle states.
The $G_0W_0$@HF and sc-$GW$ ionization energies are reported 
in panels g), h) and i) of Fig.~\ref{fig:spectrum} as a function 
of the basis set size. The $G_0W_0$
 ionization energy is calculated from 
the linearized quasi-particle equation (Eq.~\ref{eq:qpe}), 
whereas in sc-$GW$ it is extracted from the highest (valence) peak of the spectra shown in 
panels a), b) and c). 

For the quasi-particle energies corresponding to occupied states, 
the largest change is observed going from Tier 1 to Tier 2. 
For N$_2$ for example,
we observe a change in the HOMO of 
approximately $0.2$ eV going from Tier 1 (which consists of 14 NAO basis functions per atom)
to Tier 2 (39 NAO per atom).
A further increase of the basis set from Tier 2 to Tier 3 (55 NAO per atom) leads to changes 
of the order of 
$5$ meV in the HOMO -- as illustrated in the 
right panels of Fig.~\ref{fig:spectrum}. Lower lying 
quasi-particle peaks show a similar convergence behavior as the HOMO.
H$_2$O, and NH$_3$ exhibit a qualitatively similar 
behavior.
Surprisingly, for all systems considered here, 
sc-$GW$ data converge faster with the basis set 
size than perturbative $G_0W_0$ calculations. 
In the following, we will focus on closed shell molecules, 
which in many instances do not have a stable anionic state. 
Therefore, we will focus on the spectral function of occupied states only.

\begin{figure}[t]
\includegraphics[width=0.45\textwidth]{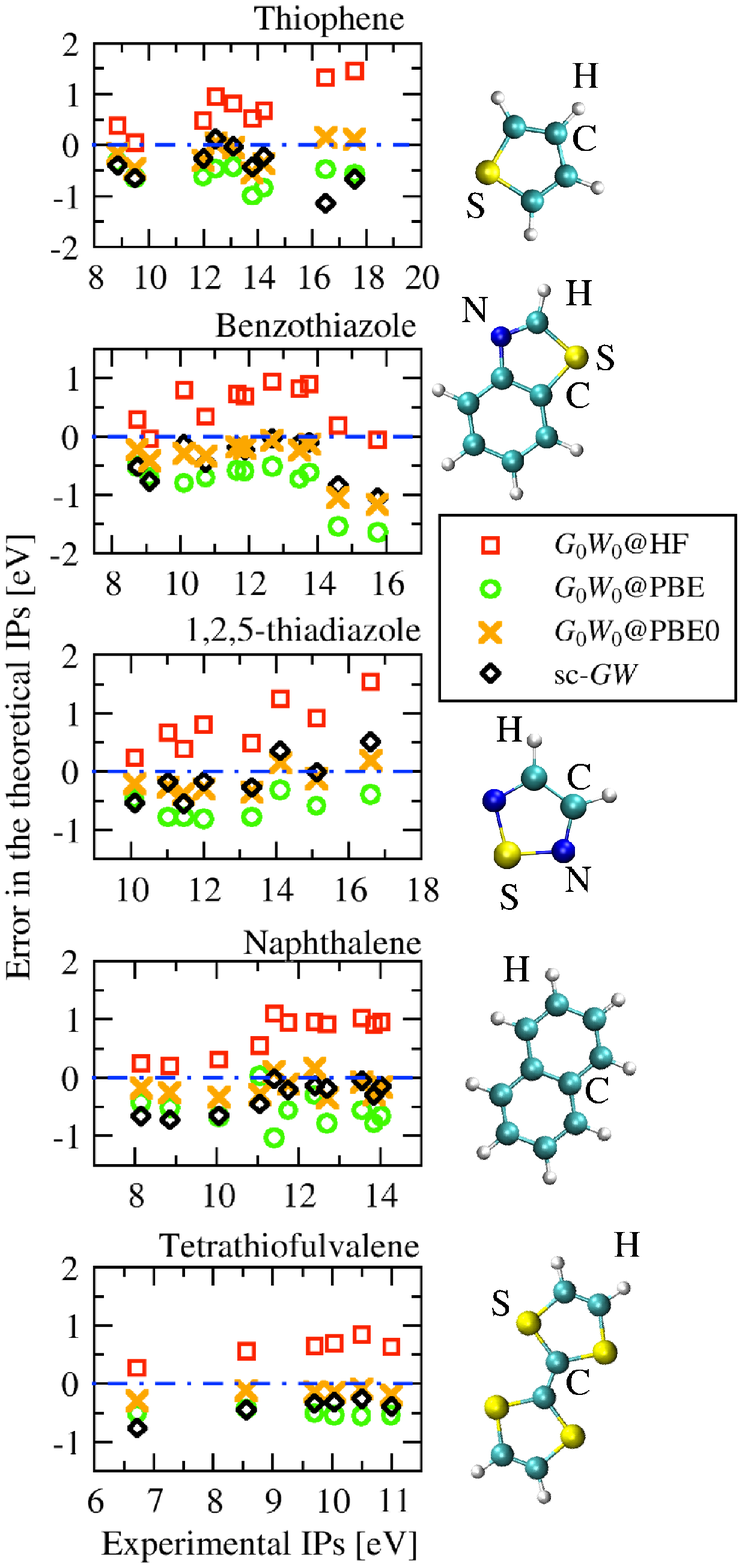}
\caption{\label{fig:trio} (Color online)
Comparison between theoretical and experimental vertical ionization energies of thiophene, 
benzothiazole, $1,2,5-$thiadiazole, naphthalene, and tetrathiafulvalene. 
Experimental photoemission data are from 
Refs.~\onlinecite{Klasinc/thiophene/1982,125Thiadiazoles/PES/2010,Benzothiazole/PES/1993,mayer:244312,19842591}.
The molecular geometries were optimized with PBE in a Tier 2 basis set and
are reported on the right.
$G_0W_0$ ionization energies are obtained with a Tier 4 basis set, the sc-$GW$ ones
with a Tier 2 basis.}
\end{figure}

To investigate the performance of the $GW$ approximation at self-consistency, 
we have performed sc-$GW$ calculations for a set of 
molecules relevant for organic photo-voltaic applications. 
We report in Fig.~\ref{fig:trio} the comparison between experimental 
\cite{Klasinc/thiophene/1982,125Thiadiazoles/PES/2010,Benzothiazole/PES/1993,mayer:244312,19842591} 
and theoretical ionization energies evaluated from sc-$GW$ 
and $G_0W_0$ based on the HF, PBE, and PBE0 starting points, for 
thiophene, benzothiazole, $1,2,5$-thiadiazole, naphthalene, and 
tetrathiafulvalene. 
For an unbiased assessment, it would be
desirable to benchmark sc-$GW$ against higher level theories, 
since in experiment the distinction between 
vertical and adiabatic ionization energies 
is difficult and vibrational effects are always present. 
For naphthalene the coupled 
cluster singles doubles with perturbative triples (CCSD(T)) method, that is currently 
considered as the gold standard in quantum chemistry, gives a vertical ionization 
potential of 8.241 eV \cite{Deleuze/etal:2003}, 
which sc-$GW$ underestimates (-7.48~eV).
For this molecule, the difference between the vertical 
and the adiabatic ionization potential is only 0.1~eV in CCSD(T). 
For thiophene, CCSD(T) calculations of the adiabatic ionization energy obtain 
8.888~eV \cite{Lo/Lau:2011}, in good agreement with experiment, whereas sc-$GW$ 
yields 8.45~eV. 
Zero-point vibration effects are small and 
cancel with core-correlation and relativistic effects. 
However, the authors of this study indicate 
that the geometry of the cation differs considerably 
from that of the molecule, but did not give 
values for the vertical ionization energy. 
It therefore remains an open question, by how much vertical 
and adiabatic ionization potentials differ for thiophene.
For benzothiazole, 
$1,2,5$-thiadiazole, and tetrathiafulvalene we were not able to find CCSD(T) 
calculations for the vertical ionization potential. 

Despite the tendency to underestimate the first 
ionization energy, for these systems sc-$GW$ ionization energies are
in good agreement with experiment, and give
a good overall description of the excitation spectrum:
full self-consistency leads to an average error of 0.4 
eV (with a maximum error of 1.2 eV) 
between the experimental and theoretical ionization energies, 
whereas HF- and PBE-based $G_0W_0$ differs on average by 0.7 eV 
(with a maximum error of 1.5 eV for $G_0W_0$@HF, and  1.6 eV for  $G_0W_0$@PBE).
Interestingly, $G_0W_0$@PBE0 ionization 
energies are close to the sc-$GW$ ones. Moreover, the $G_0W_0$@PBE0 spectrum
is in slightly better agreement with experiments with an average deviation 
of  0.3 eV (with a maximum error of 1.2 eV) -- as recently also reported for benzene
and the azabenzenes in Ref.~\onlinecite{benzenepaper}.

\begin{figure*}[t]
\includegraphics[width=0.8\textwidth]{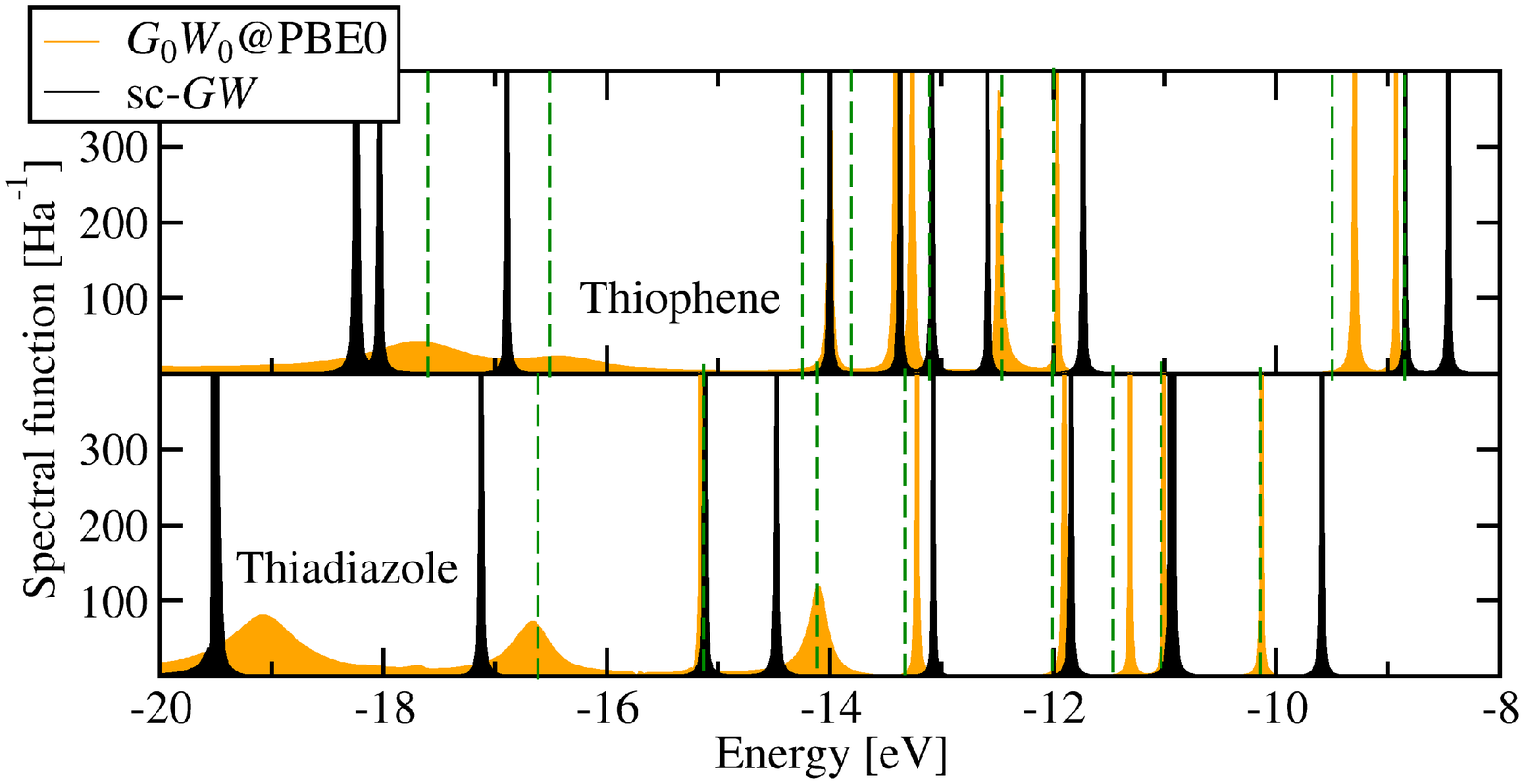}
\caption{\label{fig:trio_spectra} (Color online)
Comparison between sc-$GW$  and $G_0W_0$@PBE0 spectral functions for thiophene
and $1,2,5-$thiadiazole evaluated with a Tier 2 basis set. Vertical dashed lines indicates 
experimental photoemission data from
Refs.~\onlinecite{Klasinc/thiophene/1982} and \onlinecite{125Thiadiazoles/PES/2010},
respectively.
$G_0W_0$@PBE0 spectra are obtained with a Tier 4 basis set, the sc-$GW$ spectra
with a Tier 2 basis.
}
\end{figure*}

For small molecules, the improvements of the 
spectral properties at self-consistency can partially be traced back 
to corrections of the over- 
or under-screening in $G_0W_0$.
In PBE based $G_0W_0$ calculations, 
the small HOMO-LUMO gap induces 
an overestimation of the screening in 
the Coulomb interaction. This is the origin of
a systematic error in the $G_0W_0$@PBE 
quasi-particle energies. Similar considerations are easily 
generalized to the HF starting point, where HOMO-LUMO gaps are generally too large 
due to the missing correlation energy.  
PBE0, on the other hand, often gives a good compromise.
Therefore, the over- and under-screening is reduced in $G_0W_0$@PBE0 and the 
resulting excitation spectrum is similar to sc-$GW$, where -- because 
of the self-consistent calculation of $W$ -- this problem is mitigated. 
Based on these results, 
PBE0 emerges as an optimal 
starting point for the perturbative 
calculation of the spectral properties. 
It was argued that the screened Coulomb interaction 
obtained from sc-$GW$ may also be underscreened 
due to the lack of electron-hole interactions -- typically accounted for by 
vertex corrections \cite{PhysRevLett.94.186402}. 
This would in principle lead to a systematic 
overestimation of the quasi-particle energies, as reported 
in Ref.~\onlinecite{shishkin/vertex/kresse} for semiconductors.
Figure \ref{fig:trio}, on the other hand, indicates a slight underestimation of 
the sc-$GW$ quasi-particle energies, confirming the expectation 
that effects of the electron-hole attraction on the screened Coulomb interaction are small in molecules with 
large HOMO-LUMO gaps.

As alluded to in the introduction, previous 
sc-$GW$ studies have reported conflicting conclusions on 
the accuracy of the spectral 
properties.\cite{holmvonbarth1998,eguiluz1998,eguiluz2002,stan,thygesen}
Consequently, no consensus has so far been reached in this respect.
sc-$GW$ calculations 
for the homogeneous electron gas (HEG) indicated a
deterioration of the spectra as compared 
to perturbative $G_0W_0$ based on the local-density 
approximation (LDA). \cite{holmvonbarth1998} 
For the HEG, Holm and von Barth observed a
transfer of spectral weight from the plasmon satellite to 
the quasi-particle peak in self-consistent calculations.\cite{holmvonbarth1998} 
This results in a weaker plasmon peak and a broader valence band,
that worsens the agreement with photo-emission experiments for metallic sodium.

The first self-consistent calculation for real systems -- performed for 
potassium and silicon in the pseudo-potential 
approximation -- confirmed the picture outlined 
by Holm and von Barth, indicating a deterioration 
of the band width and band gap at self-consistency.\cite{eguiluz1998}
In a later work, Ku and Eguiluz attributed the origin of this failure
to the pseudo-potential approximation, emphasizing the importance of 
accounting for core-valence coupling in the 
determination of the screening \cite{eguiluz2002}.
However, several groups.\cite{PhysRevB.69.125212,Delaney/etal:2004,Friedrich/etal:2006} have questioned the
 convergence of these calculations with respect to the number of bands.
Nevertheless, these earlier studies gave the impression  that full self-consistency
deteriorates the spectral properties 
compared to perturbative $G_0W_0$, and that it is not 
recommended to perform sc-$GW$ calculations.
However, in our opinion, the scarce numerical evidence
for realistic systems is not enough to corroborate this notion. 

It was argued that the deterioration of spectra in sc-$GW$ might arise
due to the iterative construction of the polarizability $\chi$ as 
the product of two Green functions \cite{qpscgw2004}. 
This would systematically weaken the incoherent part of the Green function,
and reduce the intensity of the plasmon satellites.
For molecules however, this mechanism does not apply 
since quasi-particle peaks carry integer
spectral weight, and no plasmon satellites are observed.
For extended systems, this mechanism might effectively 
deteriorate the sc-$GW$ spectral function, as for the homogeneous electron gas. 
Nonetheless, more investigations are needed 
to provide a general and systematic assessment of sc-$GW$ for real solids.

\subsection{Lifetimes of quasi-particle peaks}\label{sec:life}
To facilitate the comparison between $G_0W_0$@PBE0 and sc-$GW$,  
we report in Fig.~\ref{fig:trio_spectra} the full sc-$GW$  and $G_0W_0$@PBE0 spectral function 
of thiophene and 1,2,5-thiadiazole.
Figure \ref{fig:trio_spectra} illustrates that even if the peak positions in
sc-$GW$  and $G_0W_0$@PBE0 are very similar, there are qualitative differences. 

We observe that quasi-particle peaks corresponding to high-energy
excitations are accompanied by a finite broadening. 
The broadening in turn, being inversely 
proportional to the lifetime of the corresponding
quasi-particle state, yields important information 
on the dynamics and damping of excitations.
In finite systems, finite lifetimes of electronic excitations are a well
known aspect that has been extensively discussed in the literature.
For single atoms, for which vibrational decay channels are not available, electronic lifetimes of holes have been measured with  photo-emission\cite{PhysRevA.64.012502} and pump-probe techniques\cite{nature2002}.
The most likely process that leads to the decay of holes is Auger recombination. 

Green's function theory is in principle 
exact, and is therefore expected to correctly account
for the lifetime (i.e., the broadening) of quasi-particle excitations,
if the exact self-energy were used.
The origin of peak broadening can easily 
be understood from the structure of the Lehmann 
representation: 
\begin{align}
G^{\sigma}({\bf r},{\bf r'},\omega)=
\sum_s \frac{ f_{s}^{\sigma}({\bf r }) f^{*\sigma}_{s}({\bf r'}) }{\omega - \varepsilon_s- i\eta}\quad, 
\end{align}
where we considered only holes for simplicity. We defined 
$\varepsilon_s \equiv E^{N-1}_s - E_0$, and $f_s$  are the Lehmann amplitudes. 
Here $E_0$ denotes the ground-state energy of the $N$-particle system, and $E^{N-1}_s$ the 
$s$-th excited state of the $N-1$ particle system.
If the hole left behind by the photo-emission process is close 
to the Fermi energy, the energies $E^{N-1}_s$ (and subsequently also $\varepsilon_s$) are discrete.
The spectral function therefore exhibits a series of $\delta$-functions. 
However, if the holes are low enough in energy, $E^{N-1}_s$ lies in the
continuum of the $N-1$ particle systems. Correspondingly,  
$\varepsilon_s$ varies continuously and gives rise to a series 
of delta peaks that are infinitely closely spaced and
merge into a single quasiparticle peak with a finite broadening.
 
For a quantitative assessment  of  lifetimes in molecules we would have to consider effects beyond $GW$, such as
the coupling to vibrations and the satisfaction of selection rules in the decay process. However, this goes beyond the purpose of the present work. We will therefore limit the following  discussion to the origin of lifetimes in molecules and briefly characterize their 
starting point dependence. 

For all molecules considered in this work,
the quasi-particle peaks close to the Fermi energy 
have a $\delta$-function-like character. 
This structure reflects the infinite lifetime 
of the quasi-particle excitation, and is due to 
absence of allowed electronic transitions that could 
annihilate the hole created in the photo-emission process.
The excitations of lower valence and core electrons, on the 
other hand, may have a finite lifetime, and therefore 
the quasi-particle peak gets broadened.
The physical origin of the lifetime is simple.
The hole created in a lower valence 
(or core) state by the photoemission process 
can in principle recombine with electrons 
close to the Fermi energy.
The energy released in this process 
has to be converted into an internal excitation of the system,
since isolated molecules cannot dissipate energy. 
If the energy released is larger than the HOMO-LUMO
gap, a particle-hole pair can be created. This opens 
up a scattering or decay channel for the hole, which therefore
acquires a lifetime.
The energy threshold for electron-hole formation is then given by 
$\Delta\equiv E_{\rm HOMO}^{\rm GS} - E_{\rm gap}^{\rm GS}$,
with $E_{\rm HOMO}^{\rm GS}$  and $E_{\rm gap}^{\rm GS}$ 
the HOMO level and the HOMO-LUMO gap of the starting point, respectively.
In other words, only quasi-particle states with an energy below $\Delta$ may decay, 
and acquire a finite broadening.
This argument is general and does not only apply to $GW$. 
What is particular to $G_0W_0$ is that
the relevant gap for this process is determined by the starting 
point; in this case the DFT functional for the ground state.

\begin{figure}
\includegraphics[width=0.48\textwidth]{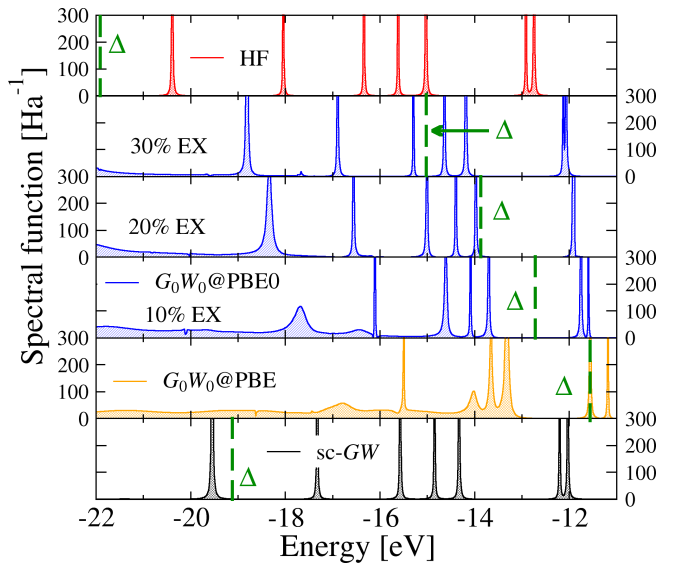}
\caption{\label{fig:lifetime} (Color online)
Spectral function of benzene evaluated from sc-$GW$ and 
$G_0W_0$ based on PBE, PBE0 with different mixture of exact exchange (EX), and HF.
Vertical dashed lines (in green) indicate the energy threshold $\Delta$  
for the formation of an electron-hole pair, which depends explicitly on the 
HOMO and LUMO levels of the underlying DFT/HF calculation. 
The quasiparticle peaks acquire a finite broadening only for states below $\Delta$.
For peaks above $\Delta$, the residual broadening stems from 
the parameter $\eta=10^{-4}$ discussed in the text. }
\end{figure}

To illustrate this effect on the 
broadening of the quasi-particle peaks, we report in Fig.~\ref{fig:lifetime}
the  spectrum of benzene evaluated from $G_0W_0$ based on different starting points 
and at self-consistency. 
Values of $\Delta$ for the different exchange-correlation functionals are 
reported as vertical dashed lines (in green).
PBE has the smallest 
HOMO-LUMO gap ($\sim5.2$ eV) and we observe a
noticeable peak broadening (i.e., short lifetime) 
at $\sim$14 eV in the $G_0W_0$@PBE0 spectrum. 
A systematic increase of the broadening is then observed  
the further a state lies below $\Delta$.
Adding exact exchange to the DFT functional and increasing its
admixture opens the HOMO-LUMO gap.  
The onset of the finite lifetime subsequently moves to lower energies.

In sc-$GW$, the broadening of the quasi-particle peaks 
is consistent with the HOMO-LUMO gap at the $GW$ level, 
and the ambiguity of the starting point dependence is lifted.  
Consequently, for benzene only peaks below $-19$ eV acquire a small finite broadening.
Based on the sc-$GW$ results, the large 
broadening observed in the $G_0W_0$@PBE0 spectrum can be attributed to the 
small HOMO-LUMO gap of the underlying PBE0 calculation. The inclusion 
of a fraction of exact exchange partially ameliorates this problem, but not fully. 
This indicates that the calculation of lifetimes presents
an additional situation in which -- due to the severe dependence on the starting point -- 
resorting to full self-consistency is essential.  

\subsection{Independence of the starting point}

Previous work showed that partially  self-consistent approaches 
such as eigenvalue self-consistent $GW$ \cite{hybertsenlouie1986} or 
quasi-particle self-consistency,\cite{qpscgw2004,qpscgw2006} 
reduce the starting point dependence but they do not eliminate 
it.\cite{benzenepaper,Liao/carter/2011}
Only full self-consistency successfully removes any dependence on the 
starting point, as we discussed in Ref.~\onlinecite{fabioprl}.
This is a major advantage of the sc-$GW$ scheme, as it allows a 
systematic assessment of the $GW$ approximation unbiased by 
spurious dependence on the input Green function. 

\begin{figure}[t]
\includegraphics[width=0.48\textwidth]{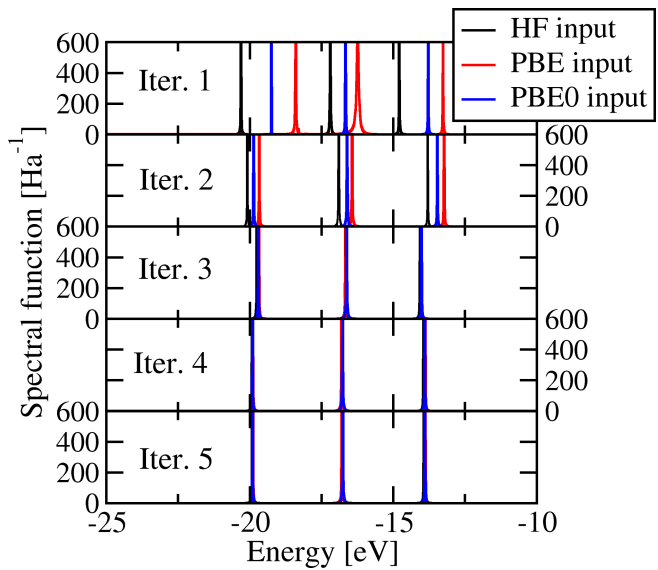}
\caption{\label{fig:uniqueness} (Color online) 
Spectral function of the CO dimer evaluated for the first five iterations of the Dyson equation 
from the HF, PBE, and hybrid PBE0 starting points. The first iteration corresponds to the $G_0W_0$ approximation.}
\end{figure}

To illustrate the independence of the starting point 
of the sc-$GW$ Green function, we report in Fig.~\ref{fig:uniqueness}
the spectral function of the carbon monoxide molecule as a function of the number of 
iterations of the Dyson equation initialized with HF, PBE, and PBE0.
After just a few iterations of the sc-$GW$ loop, the quasi-particle peaks
in the spectral function are located at the same energies demonstrating 
the independence of the starting point in sc-$GW$.

\section{Ground-state properties from the {\textit GW} approximation}\label{sec:migdal}
\subsection{Galitskii-Migdal total energy}

In MBPT, the total energy $E_{\rm tot}$ may be regarded as an explicit 
functional of the Green function, i.e. 
$E_{\rm tot}=E_{\rm tot}[G]$. 
The functional dependence of $E_{\rm tot}[G]$ on $G$ is not uniquely defined and 
different total energy functionals have been proposed over the years. 
Some example are 
the Luttinger-Ward \cite{luttingerward1960} and the Klein functional \cite{klein1961} --
which are stationary at the self-consistent Green function \cite{Almbladh/etal:1999} -- 
and the Galitskii-Migdal formula\cite{galitskii}. 
The total energy obtained from different 
functionals may differ in principle, if 
evaluated with a given Green function.
However, if the Green function is obtained self-consistently solving 
the Dyson equation, all functionals yield the same, unique total energy. 
Since we are interested in sc-$GW$ total energies, 
all total energy functionals are equivalent and
we use the Galitskii-Migdal formula because of its simplicity:
\begin{equation}
E_{\rm GM}=-{i}\int\sum_\sigma\frac{d\omega}{2\pi} 
Tr\left\lbrace\left[\omega+ {h}_0 \right] 
G^\sigma(\omega)\right\rbrace + E_{\rm ion}\label{eq:migdal} \quad.
\end{equation}
Here $h_0$ is the single-particle term of the many-body Hamiltonian, i.e. 
the sum of the 
external potential due to the nuclei 
and the kinetic energy operator,
and $E_{\rm ion}$ accounts for the repulsive nuclear energy. 
As discussed in Ref.~\onlinecite{Pablo:2002}, Eq.~\ref{eq:migdal} can be computed 
directly in imaginary frequency 
taking advantage of the reduced size of the integration grid 
needed to describe $G$ and $\Sigma$ on the imaginary axis.
In the present work, we cast Eq.~\ref{eq:migdal} 
into a more suitable form for numerical implementations\cite{DelaneyPhD}:
\begin{align}\label{eq:rewriting}
E_{\rm GM} &= -i\sum_{ij,\sigma} \overline G^\sigma_{ij}(\tau=0^-)
[2t_{ji} + 2v^{\rm ext}_{ji} +v^{\rm H}_{ji}+  \Sigma^{\rm x}_{ji,\sigma} ]\nonumber\\
 &-i \sum_{ij,\sigma} \int\frac{d\omega}{2\pi}\overline G^\sigma_{ij}(\omega)\Sigma^{\rm c}_{ji,\sigma}(\omega)e^{i\omega\eta} + E_{\rm ion} \: ,
\end{align}
where $t$ denotes the kinetic-energy operator, $v^{\rm H}$ and $v^{\rm ext}$
the Hartree and external potential, and $\Sigma^{\rm x}$ and
$\Sigma^{\rm c}$ the
exchange and correlation parts of the self-energy, respectively.
We used $\overline G \equiv s^{-1}Gs^{-1}$, with $s_{ij}=\int d{\bf r} \varphi_i({\bf r}) \varphi_j({\bf r})$.
A derivation of Eq.~\ref{eq:rewriting} is reported in Appendix \ref{sec:ap1}.
We emphasize that Eq.~\ref{eq:rewriting} is exact if evaluated with
the exact self-energy and Green function. 
In Eq.~\ref{eq:rewriting}
contributions arising from time-independent operators 
can be easily evaluated by simple 
matrix products. The correlation energy on the other hand
requires a frequency integration
which can be evaluated directly on the imaginary frequency axis without resorting to the 
analytic continuation\cite{Pablo:2002}.
We evaluated Eq.~\ref{eq:rewriting} in the $GW$ approximation.
In sc-$GW$, both $\Sigma$ and $G$ are self-consistent solution of the Dyson equation, 
whereas for $G_0W_0$ the 
self-energy is evaluated only once and $G$ is 
the non-interacting Green function of the DFT/HF calculation.
The latter procedure corresponds to a first-order perturbative correction of the DFT/HF 
total energy, with a perturbing potential given by $[\Sigma(\omega)-v_{\rm xc}]$.

\subsection{Structural parameters of diatomic molecules}

\begin{figure}[t]
\includegraphics[width=0.3\textwidth]{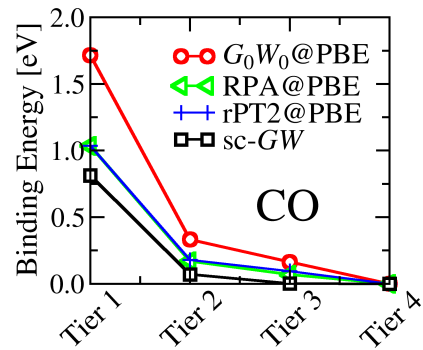}
\caption{\label{fig:bindconv} (Color online) 
BSSE corrected binding energy of CO evaluated with Tier 1, 2, 3 and 4. 
Each curve is aligned at their respective Tier 4 value. 
All data are in eV.} 
\end{figure}

Total energy differences are more important than absolute total energies, as 
they give information on structural properties and on the 
strength of chemical bonds.
Here, we restrict the discussion to the ground-state properties of dimers 
at their equilibrium geometry. The reader is referred to Ref.~\onlinecite{gw-rpa-paper}
for an assessment of the sc-$GW$ method in the dissociation limit. 

In the following, we provide an assessment of the sc-$GW$ method for 
bond lengths, binding energies, and vibrational frequencies
based on the calculations of the potential energy curve of LiH, LiF, HF, CO, H$_2$, and N$_2$.
To illustrate the convergence with the basis set,  
we report in Fig.~\ref{fig:bindconv} the binding energy of the carbon monoxide dimer 
evaluated with sc-$GW$, RPA@PBE, $G_0W_0$@PBE, and PBE-based renormalized second-order 
perturbation theory\cite{rpareview,ren/tobe} (rPT2) using increasingly larger NAO basis sets (Tier 1-4).

The mean absolute errors of 
theoretical bond lengths, binding energies, and vibrational frequencies
as compared to experiment are reported in Fig.~\ref{fig:maebond}. 
The corresponding numerical values are reported in Appendix \ref{app:data}.
\begin{figure}[t]
\includegraphics[width=0.45\textwidth]{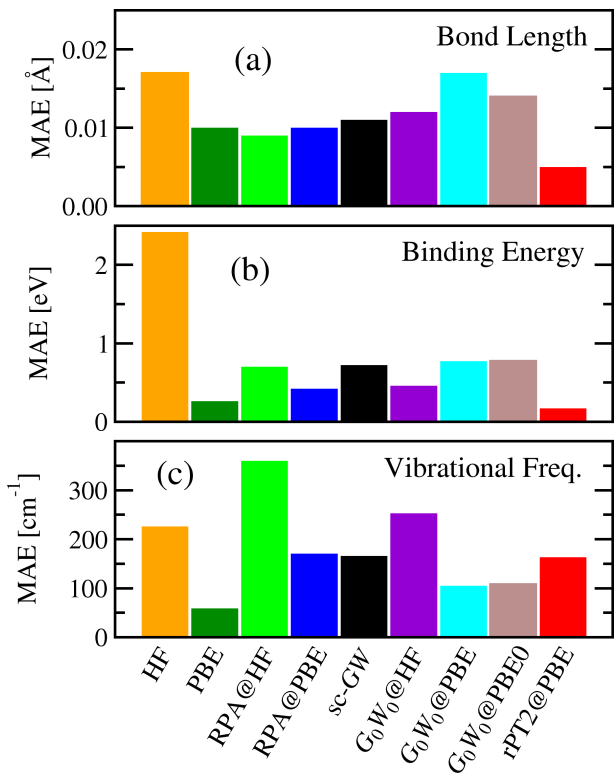}
\caption{\label{fig:maebond} (Color online) Mean absolute error (MAE) of bond lengths (upper panel),
binding energies (central panel), and vibrational frequencies (lower panel) 
of LiH, LiF, HF, CO,  H$_2$, and N$_2$ evaluated at 
different levels of theory. 
The estimated zero-point motion correction has been subtracted from the 
experimental binding energies (reported from Ref.~\onlinecite{feller:8384}).
Calculations were done with a Tier 3 basis set 
(the largest basis set available) for  H$_2$, LiH, and HF,
whereas a Tier 4 basis  was used for N$_2$ and CO.
Numerical values are reported in Tables \ref{tab:lengths},  
\ref{tab:ebinding}, and \ref{tab:vibs} in the Appendix.
}
\end{figure}
Since our calculations are performed in the Born-Oppenheimer approximation
with clamped nuclei, we compared 
our calculations with zero-point motion corrected 
experimental binding energies from Ref.~\onlinecite{feller:8384}.  
PBE, HF, and several perturbative approaches based on MBPT, namely $G_0W_0$, EX+cRPA, and rPT2
are included for comparison. 

Self-consistency provides 
better bond lengths than perturbative $G_0W_0$ calculations. 
However, the accuracy achieved by sc-$GW$ for the bond 
lengths is still comparable to perturbative RPA  
and not as good as rPT2@PBE, which 
includes higher order exchange and correlation diagrams.

The binding energies obtained from $G_0W_0$ based on HF and PBE, reported in Table \ref{tab:ebinding}, 
are systematically 
overestimated. 
Self-consistent $GW$ over-corrects this trend and yields 
binding energies that slightly underestimate experiment. 
RPA@HF and sc-$GW$ give a very similar description 
of the binding energy, the deviation between the two methods being approximately $10-20$ meV.
This similarity is expected for two reasons:
First, in diatomic molecules 
screening is small, thus the sc-$GW$ Green function resembles the Hartree-Fock one
(since in absence of polarization $W$ reduces to the bare Coulomb interaction).
Second, the RPA total energy is a  
variational functional of the Green function, and therefore RPA total energies are 
close to sc-$GW$ ones, if the input Green function is
{\it close enough} to the sc-$GW$ Green function\cite{dahlenleeuwen2006}.
Larger discrepancies between perturbative RPA and sc-$GW$ are to be
expected for the structural properties of systems 
for which the sc-$GW$ density is substantially different 
as compared to HF or semi-local DFT. 
Example of these material are molecular interfaces 
and charge transfer compounds, where the 
ground-state density (and the charge transfer) depends strongly 
on the level alignment between the 
individual components of the system.
This will be addressed in future works. 
Also, for binding energies and bond lengths, sc-$GW$ is outperformed by rPT2@PBE, which illustrates the 
importance of including exchange and correlation diagrams beyond the $GW$ approximation 
for a systematic improvement of the ground-state properties of finite systems.

For vibrational frequencies, the dependence on the starting point   is larger than for binding energies or bond lengths.
In this case, the best agreement with experiment is achieved with the PBE functional, 
whereas for HF the errors are substantially larger.
Similarly, PBE-based $G_0W_0$  and EX+cRPA 
deviate less from experiment than HF-based schemes. 
For instance, the mean absolute error of EX+cRPA@HF is approximately 
a factor of two larger that EX+cRPA@PBE, and the same is observed for $G_0W_0$.
Consequently, sc-$GW$ gives smaller MAEs compared to HF-based schemes, 
but does not improve over PBE-based perturbative schemes.

\subsection{Density and dipole moments at self-consistency}

In perturbative approaches, such as $G_0W_0$, EX+cRPA, and rPT2,
the electron density of a system is defined by the eigenstates 
of the unperturbed reference Hamiltonian -- although in principle perturbative corrections
to the eigenstates of the unpertubed Hamiltonian could be calculated. 
This constitutes a major drawback, as it is in part 
responsible for the well-known starting point dependence of perturbative schemes.
\begin{figure}
\includegraphics[width=0.35\textwidth]{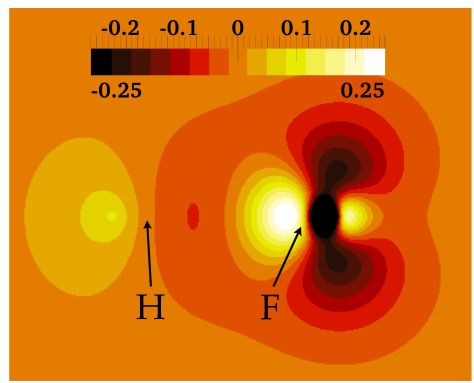}
\caption{\label{fig:dens} (Color online) Difference  
between the sc-$GW$ and Hartree-Fock
densities for the hydrogen fluoride dimer at its experimental equilibrium geometry ($d=0.917$ \AA). 
At self-consistency, electron density is shifted from negative (dark)
to positive regions (light).
Units are {\AA}$^{-3}$ and the calculation were performed
using a Tier 3 basis set.
}
\end{figure}
Self-consistency, on the other hand, permits us
to incorporate exact exchange and dynamical correlation effects 
into the electron density. 
To illustrate this aspect in sc-$GW$, we discuss in the following 
the effects of self-consistency on the density and the dipole moment of diatomic molecules.

Figure \ref{fig:dens} shows 
the density difference between 
sc-$GW$ and HF for the hydrogen fluoride dimer. 
The density difference illustrates the effects of 
$GW$ correlation on the electron density. 
sc-$GW$ enhances the angular 
distribution of the electron density exhibiting more pronounced angular correlation. 
Moreover, density is shifted from the two lobes perpendicular to the 
molecular axis to the bond region, leading to a reduction 
of the dipole moment as compared to HF, 
which is in better agreement with experiment (see Table \ref{tab:dipole}). 

The dipole moment provides a systematic way to  quantify the 
quality of the electron density of a system, as it is directly 
comparable with experimental data. 
We report in Table
\ref{tab:dipole} the dipole moment  
of LiH, LiF, HF and CO evaluated from sc-$GW$, PBE, and HF. 
sc-$GW$ dipole moments are in good agreement with 
experiments and reduce the deviation from experiment by approximately 
a factor of two compared to HF and PBE, 
that tend to under and overestimate, respectively.
The quality of the dipole moment for the small set of molecules presented here,  indicates 
that sc-$GW$ is a promising method for the 
description of charge-trasfer compounds, such as
molecular interfaces and hetero-structures. 

\begin{table}
\setlength{\tabcolsep}{10pt}
\caption{
Comparison between experimental \cite{expmol} and theoretical 
dipole moments evaluated from sc-$GW$, PBE, and HF at their 
corresponding equilibrium bond lengths.
All values are in Debye.
}
\begin{tabular}{lccccc}
\hline\hline
	 &  LiH   &	HF   & LiF	& CO    &  MAE  \\
\hline                                 
Exp. 	 &  5.88 &	1.82 & 6.28    & 0.11  &  -    \\
sc-$GW$	 &  5.90 &	1.85 & 6.48& 0.07 &  0.07 \\
PBE	 &  5.63 &	1.77 & 6.12    & 0.20  &  0.14 \\
PBE0	 &  5.77 &      1.81 & 6.20    & 0.09  &  0.06    \\
HF	 &  6.04 &      1.89 & 6.46    &-0.13  &  0.17 \\
\hline
\hline
\end{tabular}
\label{tab:dipole}
\end{table}


\section{Conclusions}\label{sec:conclusion}

We have presented an all-electron implementation of the {\it fully} 
self-consistent $GW$ method based on a numeric atom-centered orbital basis
in the FHI-aims code \cite{blum}. 
Self-consistent $GW$ is based on the iterative solution of Hedin's equations 
with the $GW$ self-energy and polarizability, and therefore it
is significantly different from partially self-consistent 
approaches based on perturbation theory,  
such as the quasi-particle self-consistent $GW$ scheme \cite{qpscgw2004,qpscgw2006} 
and self-consistency in the eigenvalues \cite{hybertsenlouie1986,kresse2007}.

In our implementation,  
the two-particle operators are treated efficiently by means 
of the resolution of identity 
technique\cite{Xinguo/implem_full_author_list}. 
The introduction of an auxiliary basis for the representation of the 
frequency and time dependence of dynamic operators, 
facilitates an accurate evaluation of 
Fourier integrals that require  just few tens of grid points.
These ingredients allow us to reformulate
Hedin's equations in a matrix form, 
that can be solved with standard linear algebra packages. 

We presented an assessment of the spectral properties of five molecules of interest for organic
photo-voltaic applications: thiophene, benzothiazole, 1,2,5-thiadiazole, naphthalene, and
tetrathiafulvalene. 
For these systems, the quasi-particle energies extracted from the sc-$GW$ spectral function are found to be in 
good agreement with experimental photo-emission data for all valence states. 
The sc-$GW$  excitation spectrum systematically improves over perturbative  $G_0W_0$
based on semi-local DFT and Hartree-Fock. 
This is interpreted as a consequence of the mitigation of 
over- and under-screening errors characteristic of  $G_0W_0$@PBE and 
$G_0W_0$@HF, respectively. 
$G_0W_0$ based on PBE0, on the other hand, provides results in slightly better agreement with experimental data than sc-$GW$. Thus, the PBE0 functional appears to be close to an optimal starting point for perturbative calculations.

Self-consistent $GW$ total energies based on the Galitskii-Migdal formula 
permit an assessment of ground-state and 
structural properties of molecules.
For a small set of diatomic molecules we evaluated binding energies, bond lengths and vibrational frequencies.
The bond lengths improve at self-consistency, but still have an accuracy comparable to other 
perturbative methods such as exact-exchange with correlation from the random-phase approximation. 
Binding energies are typically underestimated compared to experimental reference data  
and do not substantially improve over perturbative $G_0W_0$ calculations.
Our results indicate and quantify the importance of including 
vertex corrections -- or alternatively, higher order correlation and exchange diagrams -- 
in order to achieve an accurate description of the structural properties of molecules.

Finally, the dipole moments of a set of hetero-atomic dimers were studied to investigate the
accuracy of the sc-$GW$ density.
Compared to Hartree-Fock and PBE, the sc-$GW$ dipole 
moments are found in better agreement with experiment. 
These results indicate that sc-$GW$ is a promising approach for the 
description of charge-transfer compounds and hetero-junctions, where the relative ordering of the 
frontier orbitals influences the charge transfer at the interface. 

\begin{acknowledgments}
We would like to thank  Christoph Friedrich for fruitful discussions.
AR acknowledges financial support from the European Research Council Advanced Grant DYNamo (ERC-2010-AdG-267374), Spanish Grants (2010-21282-C02-01 and PIB2010US-00652), Grupos Consolidados UPV/EHU del Gobierno Vasco (IT578-13) and European Commission projects CRONOS (Grant number 280879-2 CRONOS CP-FP7).
\end{acknowledgments}

\appendix

\section{Rewriting the Galitskii-Migdal formula}\label{sec:ap1}

In Hartree atomic units ($\hbar = e = m = 1$), the electronic contribution of the 
Galitskii-Migdal total energy is \cite{fetter}:
\begin{equation}
E_{\rm GM}=-{i} \sum_\sigma \int d{\bf r} \,d t 
\lim_{\substack {{\bf r'}\rightarrow {\bf r} \\ {t' \rightarrow t^+} }} 
\left[i\frac{\partial }{\partial t} + h_0
\right] \label{eq:mig1}
G^\sigma({\bf r} t,{\bf r'}t')\quad.
\end{equation}%
Here, $h_0$ is the single-particle term of the many-body Hamiltonian, i.e., the sum of the 
kinetic energy operator and the external potential due to the nuclei.
Introducing the equation of motion for the interacting Green's function 
(see e.g. Ref.~\onlinecite{strinati/1988})
\begin{align}
&\left[i \frac{\partial }{\partial t} + \frac{\nabla^2_{\bf r}}{2} - v_{\rm H} ({\bf r}) \nonumber
-  v_{\rm ext}({\bf r}) \right]G^\sigma({\bf r} t,{\bf r'}t') - \\  \nonumber
&-\int d{\bf r''}  dt'' \Sigma^\sigma ({\bf r}t, {\bf r''}t'' ) G^\sigma ({\bf r''}t'', {\bf r'}t')= \\ 
&= \delta({\bf r}-{\bf r'}) \delta(t-t')\quad , \label{eq:motion} 
\end{align}
Eq.~\ref{eq:mig1} can be simplified by eliminating 
the partial derivative with respect to time, obtaining:
\begin{align}
E_{\rm GM}
& =-{i} \sum_\sigma\int d {\bf r}\, d {\bf r'}  dt\, dt' \nonumber \\
&\lim_{\substack {{\bf r'}\rightarrow {\bf r} \\ {t' \rightarrow t^+} }}  \nonumber
\left[ \left( -{ \nabla_{\bf r}^2} +2 v_{\rm ext}({\bf r} ) 
+  v_{\rm H}({\bf r}) \right) \delta({\bf r}-{\bf r'})\delta(t-t')\nonumber\right. \\ 
&+\left.\Sigma^\sigma ({\bf r}t, {\bf r'}t' ) 
\right] 
G^\sigma({\bf r'} t',{\bf r'}t')  \label{eq:mig3} .
\end{align}
Making use of the matrix representation of the Green function 
$G^\sigma({\bf r}, {\bf r'}, \tau) = \sum_{ij} \varphi_i({\bf r}) \overline G^\sigma_{ij}(\tau) \varphi_j({\bf r'})$ 
-- with $\overline G^\sigma \equiv s^{-1}G^\sigma s^{-1}$ --
the first three terms in Eq.~\ref{eq:mig3} can be rewritten as:
\begin{align}
 -{i} &\sum_\sigma\int d {\bf r} \lim_{{\bf r'}\rightarrow {\bf r}}  
\left[ - { \nabla_{\bf r}^2} +2 v_{\rm ext}({\bf r} )
 +  v_{\rm H}({\bf r}) \right] \times \nonumber \\
\times &\sum_{ij} \varphi_i({\bf r}) \overline G^\sigma_{ij}(\tau=0^-) \varphi_j({\bf r'})=\nonumber \\ 
 -{i} &\sum_\sigma\sum_{ij} \overline G^\sigma_{ij}(\tau=0^-)
\left[2 t_{ji} + 2 v_{ji}^{\rm ext} + v_{ji}^{\rm H} \right ]\quad. \label{eq:mig4}
\end{align}
For time independent Hamiltonians, the Green function 
depends only on time differences $\tau\equiv t-t'$. 
In the last step of Eq.~\ref{eq:mig4}, we defined
the matrix representation of the kinetic energy operator as 
$t_{ij} = \int d {\bf r} \varphi_i({\bf r}) \left[- \frac{ \nabla_{\bf r}^2}{2}\right]  \varphi_j({\bf r}) $,
and use a similar representation for $ v_{ji}^{\rm ext}$  and $ v_{ji}^{\rm H}$.
Finally, the last term in Eq.~\ref{eq:mig3} can be rearranged by using the Fourier transform
of the Green function and the self-energy
$G( t, t') = \int_{-\infty}^{+\infty} \frac{d\omega}{2\pi} e^{-i\omega (t-t') } G(\omega)$, and 
substituting the matrix representation of $G$:
\begin{align}
&-{i}\sum_\sigma \int d {\bf r}\, d {\bf r''}  dt\, dt'' 
\lim_{\substack {{\bf r'}\rightarrow {\bf r} \\ {t' \rightarrow t^+} }} 
\Sigma^\sigma ({\bf r}t, {\bf r''}t'' ) 
G^\sigma({\bf r''} t'',{\bf r'}t') = \nonumber \\ 
&-{i} \sum_\sigma\sum_{ij} \int \frac{d\omega}{2\pi}
\Sigma^\sigma_{ji} (\omega) 
 G^\sigma_{ij} (\omega) e^{i\omega\eta}\quad.
\label{eq:term2}  
\end{align}
Summing Eqs.~\ref{eq:mig4} and \ref{eq:term2} and separating the 
self-energy in its exchange and correlation components
$\Sigma_{ij}(\omega) = \Sigma_{ij}^{\rm x} + \Sigma_{ij}^{\rm c} (\omega)$, one finally 
arrives at the expression reported in Eq.~\ref{eq:rewriting}.
We refer to Ref.~\onlinecite{Pablo:2001}, for a discussion on the evaluation of the Galitskii-Migdal formula 
on the imaginary frequency axis.

\section{Binding energies, bond lengths, and vibrational frequencies of diatomic molecules}\label{app:data}
In Tables \ref{tab:ebinding} and \ref{tab:lengths} we report
counterpoise corrected\cite{liu/BSSE/1973} binding energies and bond lengths
for H$_2$, LiH, LiF, HF, N$_2$, and CO.
The corresponding vibrational frequency are reported in Table \ref{tab:vibs}.
The sc-$GW$ results are compared with experimental values\cite{feller:8384,expmol} and 
several perturbative approaches based on MBPT, namely $G_0W_0$, EX+cRPA, and rPT2@PBE\cite{rpareview}.
HF and PBE  are included for comparison.


\begin{table*}[th]
\setlength{\tabcolsep}{4.5pt}
\caption{
sc-$GW$, and perturbative $G_0W_0$ binding energies (evaluated from the Galitskii-Migdal formula) 
of diatomic molecules compared to 
(zero point motion corrected)
experimental reference data taken from Ref.~\onlinecite{feller:8384}. 
We report perturbative RPA, HF, PBE, and renormalized second-order perturbation theory (rPT2) for 
comparison. 
Calculations were done with a Tier 3 basis set (the largest basis set available) for  H$_2$, LiH, and HF, 
whereas a Tier 4 basis  was used for N$_2$ and CO.
The mean absolute errors (MAE) 
are reported in panel (b) of Fig.~\ref{fig:maebond}.
All values are in eV.
}
\begin{tabular}{lcccccc|c}
\hline\hline
                  &    H$_2$	&LiH	&HF	&LiF	&N$_2$	&CO     & MAE\\ 
\hline                                                                    
Exp     	  &  -4.75	&-2.52	&-6.12	&-6.02	&-9.91	&-11.24 &   \\
sc-$GW$           &  -4.41	&-2.16	&-5.55	&-5.50	&-8.42	&-10.19 & 0.72  \\
$G_0W_0$@HF       &  -5.05	&-2.72	&-6.45	&-6.60	&-10.61	&-11.88 & 0.46  \\
$G_0W_0$@PBE      &  -5.44	&-2.94	&-6.46	&-6.37	&-11.82	&-12.16 & 0.77  \\
$G_0W_0$@PBE0     &  -5.32      &-2.90  &-6.49  &-6.67  &-11.50 &-12.34 & 0.78  \\
RPA@HF            &  -4.41	&-2.17	&-5.54	&-5.52	&-8.51	&-10.19 & 0.70  \\
RPA@PBE           &  -4.68	&-2.32	&-5.60	&-5.43	&-9.54	&-10.48 & 0.42  \\
rPT2@PBE          &  -4.71	&-2.49	&-5.93	&-5.90	&-9.42	&-11.06 & 0.72  \\
HF                &  -3.64	&-1.49	&-4.22	&-3.95	&-5.10	&-7.62  & 2.42  \\
PBE               &  -4.54	&-2.32	&-6.17	&-6.03	&-10.58	&-11.67 & 0.26  \\
\hline
\hline
\label{tab:ebinding}
\end{tabular}
\end{table*}
%
\begin{table*}[th]
\setlength{\tabcolsep}{4.5pt}
\caption{
sc-$GW$ and perturbative $G_0W_0$ bond lengths of diatomic molecules compared to 
experimental reference data taken from Ref.~\onlinecite{expmol}. 
RPA, HF, PBE, and renormalized second-order perturbation theory (rPT2) are included for 
comparison. 
Mean absolute errors (MAE) are reported 
in panel (a) of Fig.~\ref{fig:maebond}. 
All values are in \AA.
}
\begin{tabular}{lcccccc|c}
\hline\hline
              &  H$_2$  &   LiH &    HF   &    LiF   & N$_2$&  CO   & MAE  \\
\hline                                                                
Exp.          &  0.741  &  1.595&   0.917 &    1.564 & 1.098&  1.128  &   \\
sc-$GW$       &  0.735  &  1.579&   0.919 &    1.586 & 1.085&  1.118  & 0.011 \\
$G_0W_0$@HF   &  0.733  &  1.560&   0.919 &    1.579 & 1.093&  1.119  & 0.012 \\
$G_0W_0$@PBE  &  0.746  &  1.582&   0.938 &    1.593 & 1.116&  1.143  & 0.017 \\
$G_0W_0$@PBE0 &  0.741	&  1.564&   0.932 &    1.590 & 1.100&  1.136  & 0.014 \\
RPA@HF        &  0.734  &  1.587&   0.914 &    1.576 & 1.087&  1.117  & 0.009 \\
RPA@PBE       &  0.745  &  1.597&   0.927 &    1.589 & 1.107&  1.137  & 0.010 \\
rPT2@PBE      &  0.739  &  1.597&   0.914 &    1.578 & 1.091&  1.125  & 0.005 \\
HF            &  0.734  &  1.606&   0.898 &    1.560 & 1.066&  1.102  & 0.017 \\
PBE           &  0.751  &  1.605&   0.930 &    1.575 & 1.104&  1.136  & 0.010 \\
\hline
\hline
\label{tab:lengths}
\end{tabular}
\end{table*}

\begin{table*}[th]
\setlength{\tabcolsep}{4.5pt}
\caption{
sc-$GW$, and perturbative $G_0W_0$ vibrational frequencies of diatomic molecules compared to 
experimental reference data taken from Ref.~\onlinecite{expmol}. 
RPA, HF, PBE, and renormalized second-order perturbation theory (rPT2) 
are included for comparison. 
The mean absolute errors (MAE) are reported 
in panel (c) of Fig.~\ref{fig:maebond}. 
All values are in cm$^{-1}$.
}
\begin{tabular}{lcccccc|c}
\hline\hline
              &  H$_2$  &   LiH &    HF   &    LiF   & N$_2$&  CO & MAE    \\
\hline                                                                
Exp.          & 4401 &  1405 &  4138 &  911   & 2359 &  2170  &  \\
sc-$GW$       & 4533 &  1743 &  4266 &  971   & 2543 &  2322  & 166 \\
$G_0W_0$@HF   & 4585 &  1827 &  4341 &  1010  & 2490 &  2647  & 252  \\
$G_0W_0$@PBE  & 4341 &  1743 &  4130 &  971   & 2346 &  2322  & 105  \\
$G_0W_0$@PBE0 & 4425 &  1813 &  4273 &  922   & 2386 &  2222  & 109  \\
RPA@HF        & 4533 &  1685 &  5512 &  952   & 2544 &  2321  & 360  \\
RPA@PBE       & 4357 &  1691 &  4757 &  933   & 2354 &  2115  & 172 \\
HF            & 4567 &  1473 &  4569 &  949   & 2736 &  2448  & 226 \\
PBE           & 4320 &  1364 &  3991 &  899   & 2328 &  2128  & 59  \\
rPT2@PBE      & 4460 &  1605 &  4620 &  922   & 2507 &  2251  & 163 \\

\hline
\hline
\label{tab:vibs}
\end{tabular}
\end{table*}

\newpage

%

\end{document}